\documentclass[12pt]{article}

\usepackage{latexsym,amsmath,amscd,amssymb,graphics,cite,svg}
\usepackage{authblk}
\usepackage{enumerate}
\usepackage{cancel} 
\usepackage{graphicx}
\usepackage{framed}
\usepackage[colorlinks]{hyperref}
\usepackage{url}
\usepackage{cite}
\usepackage[margin=1in]{geometry}
\usepackage{lipsum}
\usepackage{graphicx,caption}

\usepackage[all]{xy}

\newtheorem{theorem}{Theorem}

\newtheorem{lemma}[theorem]{Lemma}

\newtheorem{remark}[theorem]{Remark}

\def\be{\begin{equation}}
\def\ee{\end{equation}}
\def\bea{\begin{eqnarray}}
\def\eea{\end{eqnarray}}
\def\ba{\begin{array}}
\def\ea{\end{array}}

\def\boldeta{\boldsymbol{\eta}}

\let\<\langle
\let \>\rangle

\newcommand{\rem}[1]{}

\newcommand{\de}{\delta}

\newcommand{\bvarphi}{\boldsymbol{\varphi}}

\newcommand{\bu}{\boldsymbol{u}}

\newcommand{\bX}{\boldsymbol{X}}

\newcommand{\bF}{\boldsymbol{F}}

\newcommand{\bY}{\boldsymbol{Y}}

\newcommand{\bn}{\mathbf{n}}

\newcommand{\pp}[2]{\frac{\partial #1}{\partial #2}}

\newcommand{\dede}[2]{\frac{\delta #1}{\delta #2}}

\usepackage{framed}

\newcommand{\todo}[1]{\vspace{5 mm}\par \noindent
\framebox{\begin{minipage}[c]{0.95 \textwidth}
\tt #1 \end{minipage}}\vspace{5 mm}\par}

\begin{document}

\title{Variational geometric approach to the thermodynamics of porous media}
\author{Fran\c{c}ois Gay-Balmaz$^{1}$, Vakhtang Putkaradze$^{2,3}$}
\addtocounter{footnote}{1}
\footnotetext{LMD - Ecole Normale Sup\'erieure de Paris - CNRS, 75005, Paris.
\texttt{francois.gay-balmaz@lmd.ens.fr}}
\addtocounter{footnote}{1}
\footnotetext{ATCO SpaceLab, 5302 Forand St SW, Calgary, AB, T3E 8B4, Canada.}
\addtocounter{footnote}{1}
\footnotetext{Department of mathematical and statistical sciences, University of Alberta, 
Edmonton, AB, T6G 2G1 Canada. \texttt{putkarad@ualberta.ca}}


\maketitle

\abstract{Many applications of porous media research involves high pressures and, correspondingly, exchange of thermal energy between the fluid and the matrix. While the system is relatively well understood for the case of non-moving porous media, the situation when the elastic matrix can move and deform, is much more complex. In this paper we derive the equations of motion for the dynamics of a deformable porous media which includes the effects of friction forces, stresses, and heat exchanges between the media, by using the new methodology of variational approach to thermodynamics \cite{gay2017lagrangian,gay2017lagrangian2}. This theory extends the recently developed variational derivation of the mechanics of deformable porous media \cite{FaFGBPu2020} to include thermodynamic processes and can easily include incompressibility constraints. The model for the combined fluid-matrix system, written in the spatial frame, is developed by introducing mechanical and additional variables describing the thermal energy part of the system, writing the action principle for the system, and using a nonlinear, nonholonomic constraint on the system deduced from the second law of thermodynamics. The resulting equations give us the general version of possible friction forces incorporating thermodynamics, Darcy-like forces and friction forces similar to those used in the Navier-Stokes equations. The equations of motion are valid for arbitrary dependence of the kinetic and potential energies on the state variables. The results of our work are relevant for geophysical applications, industrial applications involving high pressures and temperatures, food processing industry, and other situations when both thermodynamics and mechanical considerations are important.}

\maketitle


\section{Introduction} 

\subsection{Review of the literature related to porous media dynamics}

Many industrial processes require the use of porous materials in changing temperature and pressure conditions. There are numerous applications such as porous filters for nuclear industry \cite{petlitckaia2017synthesis}, geophysics \cite{mctigue1986thermoelastic,veveakis2014thermo} and gas particle filters, in particular, for use with diesel-power engines \cite{chen2011effects}. The importance of multi-scale modeling involving poromechanics for food industry has also been discussed \cite{ho2013multiscale}. In terms of thermal considerations in food industry, there is an interesting application of food change using microwave heating \cite{rakesh2011microwave}. Because of the wide range of applicability of this problem, it is useful to construct a theory with maximum universality and flexibility, capable to describe the behavior of a variety of materials for the elastic matrix and fluid, both compressible and incompressible, systematically derivable from the first principles. 

Before we move into the discussion of thermal effects, it is useful to start with a short review of earlier works in poromechanics addressing only mechanical properties of the media. Due to the large amount of work in the area, our description must necessarily be brief and only focus on the works essential to this discussion. The earlier developments in the field of poromechanics are due to K. von Terzaghi \cite{Te1943} and M. Biot \cite{biot1941general,biot1955theory,biot1957elastic} in the consolidation of porous media, and subsequent works by M. Biot which derived the time-dependent equations of motion for poromechanics, based on certain assumptions on the media. M. Biot also considered the wave propagation in both low and high wavenumber regime  \cite{biot1962mechanics,biot1962generalized,biot1963theory,biot1972theory}. The amount of recent work in the field of porous media is vast, both in the field of model development  \cite{joseph1982nonlinear,detournay1993fundamentals,dell1998micro,brovko2007continuum,carcione2010computational,grillo2014darcy} and their subsequent mathematical analysis   \cite{showalter2000diffusion,bociu2016analysis,bastide2018penalization}. We refer the reader interested in the history of the field to the  review \cite{rajagopal2007hierarchy} for a more detailed exposition of the literature. 

Biot's work still remain highly influential today, especially in the field of acoustic propagation of waves through porous media. However, subsequent careful investigations have revealed difficulties in the interpretation of various terms through the general principles of  mechanics, such as material objectivity, frequency-dependent permeability and changes of porosity in the model, as well as the need to describe large deformations of the model \cite{wilmanski2006few}. Based on this criticism, \cite{wilmanski2006few} develops an alternative approach to saturated porous media equations which does not have the limitations of the Biot's model. Alternatively, \cite{ChMo2010,ChMo2014,vuong2015general} develop equations for saturated porous media  based on the general thermodynamic principles of mechanics. While our approach is different from the two papers cited above, the resulting equations we obtain in this paper, when appropriate, compare well to the results obtained in \cite{ChMo2010,ChMo2014,vuong2015general}. 

The mainstream approach to the porous media has been to treat the dynamics as being friction-dominated by dropping the inertial terms from the equations. The seminal book of Coussy \cite{coussy1995mechanics} contains a lot of background information and analysis on that approach. For more recent work, we will refer the reader to, for example, the studies of multi-component porous media flow  \cite{seguin2019multi}, as well as the gradient approach to the thermo-poro-visco-elastic processes \cite{both2019gradient}. A thorough review of recent progress using this approach can be found in a recent detailed book \cite{cheng2016poroelasticity}. In particular, chapter 11  of this book describes thermal effects of poroelasticity in the classical approach in great details, and we refer the reader interested in the state-of-the-art of the classical approach to poroelasticity to that book.

Our work, in contrast, is dedicated to the development of modes using variational principles of mechanics, which is a sub-field of all the approaches to porous media. The equations we will derive here, without the viscous terms, will be of infinite-dimensional Hamiltonian type and approximate the inertia terms and large deformations consistently. On the other hand, the friction-dominated approach gives equations of motion that are of gradient flow type. 

Fluid-filled elastic porous media, by its very nature, is a highly complex object involving both the individual dynamics of the fluid and the media, and highly nontrivial interactions between them. It is not realistic to assume complete knowledge of the micro-structured geometry of pores in the elastic matrix and the details of fluid motion inside the pores. Hence, models of porous media must include  interactions between the macroscopic dynamics and an accurate, and yet treatable, description of \emph{relevant aspects} of the micro-structures. Because of the large variations of the geometry and dynamics of micro-structures between different porous media (e.g. biological materials vs geophysical applications) the task of deriving a detailed, unified theory of porous media suitable for all applications is most likely not possible. However, provided a detailed set of assumptions and understanding the limitations of each particular assumption, deriving a consistent theory is possible. In such a framework, we believe, the variational theory is advantageous since it can develop a consistent mathematical model once the physical assumptions on the system are formulated and the relevant terms identified. Variational methods are developed by first describing the Lagrangian of the system on an appropriate configuration manifold, and then computing the critical curves of the associated action functional to obtain the equations of motion in a systematic way. The advantage of the variational methods is their consistency, as opposed to approaches based on balancing the conservation laws for a given point, or volume, of fluid. In a highly complex system like poromechanics, especially when written in the non-inertial Lagrangian frame associated with the matrix, writing out all the forces and torques to obtain correct equations is very difficult. In contrast, the equations of motion  follow from variational methods automatically, and the conservation laws are obtained in a general setting, \emph{i.e.} for arbitrary Lagrangians, as long as necessary symmetry arguments are satisfied.

 One of the earliest papers in the field utilizing variational methods was \cite{bedford1979variational} where the kinetic energy of pore expansion was incorporated into the Lagrangian to obtain the equations of motion. In that work, several Lagrange multipliers were introduced to enforce the continuity equation for both solid and fluid.
The works 
\cite{aulisa2007variational,aulisa2010geometric} use variational principles for explanation of the Darcy-Forchheimer law. Furthermore, 
\cite{lopatnikov2004macroscopic,lopatnikov2010poroelasticity} derive the equations of porous media using additional terms in the Lagrangian coming from the kinetic energy of the microscopic fluctuations. 
Of particular interest to us are the works on the Variational Macroscopic Theory of Porous Media (VMTPM) which was formulated in its present form in 
\cite{dell2000variational,sciarra2008variational,madeo2008variational,dell2009boundary,serpieri2011formulation,serpieri2015variationally,serpieri2016general,auffray2015analytical,serpieri2016variational,travascio2017analysis}, also summarized in a recent book \cite{serpieri2017variational}. In these works, the microscopic dynamics of capillary pores is modelled by a second grade material, where the internal energy of the fluid depends on both the deformation gradient of the elastic media, and the gradients of local fluid content.  The study of a pre-stressed system using variational principles and subsequent study of propagation of sound waves was undertaken in \cite{placidi2008variational}.

One of the main assumptions of the VMTPM is the dependence of the internal energy of the fluid on the quantity measuring the micro-strain of the fluid, or, alternatively, the fluid content or local density of fluid, including, in some works, the gradients of that quantity. This assumption is physically relevant for compressible fluid, but, in our view, for an incompressible fluid (which, undoubtedly, is a mathematical abstraction), such dependence is difficult to interpret. For example, for geophysical applications, fluids are usually considered compressible because of the large pressures involved. In contrast, for biological applications like the dynamics of highly porous sponges in water, the compressibility effects of the water, and, as we shall discuss here, of the sponge itself, can be neglected. For a truly incompressible fluid, it is difficult to assign a physical meaning to the dependence of internal energy of the fluid on the parameters of the porous media, as noted by \cite{wilmanski2006few}.  

This origin of this difficulty in the interpretation of fluid content was explained in \cite{FaFGBPu2020}, in terms of the fluid content being a constraint in the fluid's incompressibility, and the fluid pressure being the Lagrange multiplier related to the incompressibility.  That description in \cite{FaFGBPu2020} was based on the classical Arnold description of incompressible fluid as geodesic motion, hence Euler-Lagrange equations, on the group of volume-preserving diffeomorphisms of the fluid domain \cite{arnold1966geometrie}. In Arnold's theory the Lagrangian is simply the kinetic energy, as the potential energy of the fluid is absent, and the fluid pressure enters the equations from the incompressibility condition. This paper extends the initial derivation of \cite{FaFGBPu2020,FaFGBPu2020_2}, to include thermodynamic processes via the variational approach to thermodynamics initially developed in \cite{gay2017lagrangian,gay2017lagrangian2}. In particular, we achieve the following novel results:
\begin{enumerate} 
\item We derive a consistent theory of two-phase porous media with thermal effects for arbitrary Lagrangians and equations of state for both the fluid and the solid.
\item We show how to incorporate incompressibility constraints in the variational principle of thermodynamics.
\item We derive the equations for the cases when either the fluid and/or the solid can be compressible and incompressible.
\item We derive the balance of energy for each media and for the total system, which illustrates the effects of the irreversible processes.
\item We explain how the variational thermodynamics approach allows us to formulate the general form of dissipative terms for forces and stresses. In particular, we derive the explicit functional dependence on the inverse temperature that is necessary for thermodynamic consistency and, to our knowledge, has not been discovered before.  
\end{enumerate}


It is also useful to have a short discussion on the choice of coordinates and physics of what is commonly considered the saturated porous media. In most, if not all, previous works, undertaken by authors other than us, the saturated porous media is a combined object consisting of an (elastic) dense matrix, and a network of small connected pores filled with fluid. The fluid encounters substantial resistance when moving through the pores due to viscosity and the no-slip condition on the boundary. In such a formulation, it is easier to consider the motion of the porous matrix to be 'primary', and the motion of the fluid to be computed with respect to the porous matrix itself. Because the motion of the elastic matrix is 'primary', the equations are written in the system of coordinates consistent with the description of the elastic media, which is the material frame associated with the media. In this paper, we take an alternative view, namely, we choose the same coordinate system of a stationary observer (Eulerian frame) for the description of both the fluid and the elastic media. This is the approach taken in \cite{FaFGBPu2020,FaFGBPu2020_2}. This frame is more frequently used in the classical fluid description, but is less common in the description of elastic media. Ironically, the Eulerian frame is also frequently used in the description of wave propagation in the media, in particular, classical Biot's theory \cite{biot1962mechanics,biot1962generalized,biot1963theory,biot1972theory}. 
Physically, our description is more relevant for the case of a porous media consisting of a dense network of elastic 'threads' positioned inside the fluid, which is a case that has not been considered before apart from \cite{FaFGBPu2020}. In our formulation, we choose the Eulerian description for both the fluid and the elastic matrix. It is worth noting that the combined Eulerian description is also applicable to the regular  porous media with a 'dense' matrix, and is also well suited for the description of wave propagation in such media as shown in \cite{FaFGBPu2020} comparing the results of variational models with that of Biot. 
We shall also point out that our theory can be reformulated and is applicable for the familiar choice of the Lagrangian material description with respect to the elastic porous matrix. These descriptions are completely equivalent from the mathematical point of view, and this is rigorously justified by using the process of Lagrangian reduction by symmetry in continuum mechanics \cite{GBMaRa12}.

\subsection{Lagrangian variational approach to mechanics and thermodynamics.} The variational approach that we exploit in this paper is based on the Hamilton principle of classical mechanics and on its extension to include irreversible processes developed in \cite{gay2017lagrangian,gay2017lagrangian2,GBYo2019}, which is quickly reviewed below.

Consider a mechanical system with configuration manifold $Q$ and Lagrangian
\begin{equation}\label{L_TQ}
L:TQ \rightarrow \mathbb{R}
\end{equation} 
defined on the tangent bundle of $Q$, usually given by the kinetic energy minus the potential and internal energies of the system. In absence of irreversible processes and external forces, the equations of motion for the system are given by the Euler-Lagrange equations for $L$. They arise from the Hamilton principle
\begin{equation}\label{HP_Q} 
\delta \int_0^T L(q, \dot q){\rm d} t=0
\end{equation} 
for arbitrary variations $ \delta q$ of the curve $q(t)$ with $ \delta q_{t=0,T}=0$. In Section \S\ref{subsec_review_variational} we review the extension of these ideas for the case a continuum system, with $Q$ infinite dimensional.

We now present the extension of \eqref{HP_Q} to nonequilibrium thermodynamics in its simplest case, i.e., the case of a mechanical system with only one entropy variable $S \in \mathbb{R} $, see \cite{gay2017lagrangian}. This theory is applicable to finite-dimensional mechanical models with thermodynamics, such as the problem of a piston \cite{gay2017lagrangian}. The Lagrangian is a function
\[
L:TQ \times \mathbb{R} \rightarrow \mathbb{R} 
\]
and the variational formulation consists in the variational condition
\begin{equation}\label{VaCond} 
\delta \int_0^T L(q, \dot q,S){\rm d} t=0
\end{equation} 
subject to the constraint (phenomenological constraint)
\begin{equation}\label{PhCo} 
\frac{\partial L}{\partial S} \dot S = \left\langle F^{\rm fr} , \dot q \right\rangle 
\end{equation} 
on the solution curve and the constraint (variational constraint)
\begin{equation}\label{VaCo} 
\frac{\partial L}{\partial S} \delta S = \left\langle F^{\rm fr} , \delta  q \right\rangle 
\end{equation} 
on the variations $ \delta q$ and $ \delta S$.

In \eqref{PhCo} and \eqref{VaCo}, $F^{\rm fr}:TQ \times \mathbb{R} \rightarrow T^*Q$ is the friction force describing the irreversible process of the system, $F^{\rm fr}(q, \dot q, S) \in T^*_qQ$, with $T^*Q$ the cotangent bundle to $Q$. In absence of irreversible process, i.e., with $F^{\rm fr}=0$, the variational formulation \eqref{VaCond}--\eqref{VaCo} recovers the Hamilton principle \eqref{HP_Q}. Application of \eqref{VaCond}--\eqref{VaCo} yields the system of equations
\begin{equation}\label{thermo_mech}
\left\{ 
\begin{array}{l}
\displaystyle\vspace{0.2cm}\frac{d}{dt} \frac{\partial L}{\partial\dot q}- \frac{\partial L}{\partial q}= F^{\rm fr}\\
\displaystyle \frac{\partial L}{\partial S} \dot S = \left\langle F^{\rm fr} , \dot q \right\rangle 
\end{array}
\right.
\end{equation} 
describing the thermomechanical evolution of the system. In accordance with the first law of thermodynamics for an isolate system, the total energy of the system $E_{\rm tot}= \left\langle \frac{\partial L}{\partial \dot q}, \dot q \right\rangle -L $ is preserved. The partial derivative
\[
\frac{\partial L}{\partial S}=:- T 
\]
is identified with minus the temperature of the system and it is assumed $T>0$. We thus have the entropy equation
\[
\dot S = - \frac{1}{T} \left\langle F^{\rm fr} , \dot q \right\rangle
\]
and the second law of thermodynamics imposes $F^{\rm fr}$ to be a dissipative force: $\left\langle F^{\rm fr} (q, \dot q, S), \dot q \right\rangle \leq 0$. We refer to \cite{gay2017variational,GBYo2018,GBYo2019} for the extension of \eqref{VaCond}--\eqref{VaCo} to other irreversible processes and to open systems. We will review in \S\ref{subsec_example_piston} an extension of \eqref{VaCond}--\eqref{VaCo} with heat conduction relevant for porous media thermodynamics.

We note that the variational formulation \eqref{VaCond}--\eqref{VaCo} involves two types of constraints: a \textit{kinematic} constraint \eqref{PhCo} on the solution curve of \eqref{VaCond} and a \textit{variational} constraint \eqref{VaCo} on the variations to be considered in  \eqref{VaCond}. As it is clearly apparent, the two constraints are related in a systematic way which consists in replacing the time derivatives by $\delta$-variations. This type of variational formulation with two classes of constraint is a nonlinear extension of the Lagrange-d'Alembert variational formulation for nonholonomic systems. For nonholonomic systems with linear nonholonomic constraints the kinematic and variational constraints take the form
\[
A^a(q) \cdot \dot q =0 \qquad\text{and}\qquad A^a(q) \cdot \delta q=0, \quad a=1,...,m
\]
for some vector valued one-form $A$ on $Q$. The two constraints are also related by a formal replacement of the time derivative $\dot q$ by a $\delta$-variation $ \delta q$.

\color{black} 

\section{Equations of motions for porous media with entropy considerations}
\label{sec:3D_eqs} 
In this section, we derive the equations of motion for a porous media with internal entropy, but without irreversible thermodynamics processes. More precisely, we derive the equations of motion for a porous medium filled with an incompressible fluid, for the case of a solid elastic matrix, by using a variational formulation deduced from Hamilton's principle. The derivation in this Section follows \cite{FaFGBPu2020} with the additional inclusion of entropy in the considerations. Since most of the major details of this derivation are already contained in \cite{FaFGBPu2020}, we will go through this derivation quickly to just outline the definitions and main ideas.  

For both the fluid and the elastic matrix, we shall follow the differential geometric description outlined in the book by Marsden and Hughes \cite{marsden1994mathematical}, where the reader can find the background and fill in technical details of the description of each media. We start with some necessary background information on the description of the combined dynamics of the elastic media and fluid that is contained in it.

\subsection{Geometric variational formulation in continuum mechanics}\label{subsec_review_variational}

We quickly review here the geometric variational formulation for a single continuum, which can be either a fluid or an elastic body. We assume that the motion of the continuum in $ \mathbb{R} ^3  $ is described by a time dependent smooth embedding $\boldsymbol{\varphi }  (t,\_\,): \mathcal{B} \rightarrow \mathbb{R} ^3  $ which attributes to each material point $ \boldsymbol{X}  \in \mathcal{B} $ its location
\[
\boldsymbol{x} = \boldsymbol{\varphi }  (t, \boldsymbol{X} )
\]
in $\mathbb{R} ^3  $ at time $t$. Here $ \mathcal{B} \subset \mathbb{R} ^3  $ denotes the reference configuration of the continuum, assumed to be a compact domain with smooth boundary, and $\mathcal{B} _t= \boldsymbol{\varphi }  (t, \mathcal{B} ) \subset \mathbb{R} ^3$ is the domain occupied by the continuum at time $t$.

\subsubsection{Lagrangians}

In absence of irreversible processes, the dynamics of the continuum is completely described by the time dependent embedding $ \boldsymbol{\varphi }  (t,\_\,)$ hence its configuration space is the infinite dimensional manifold $ Q= \operatorname{Emb}(\mathcal{B} , \mathbb{R} ^3  ) $ of all smooth embeddings of $ \mathcal{B} $ into $ \mathbb{R} ^3$ and the Lagrangian is a map
\[
L:T \operatorname{Emb}( \mathcal{B} , \mathbb{R} ^3  ) \rightarrow \mathbb{R} 
\]
defined on the tangent bundle of $ \operatorname{Emb}( \mathcal{B} , \mathbb{R} ^3  )$, exactly as in \eqref{L_TQ}, given by the kinetic minus the internal and potential energies of the continuum. For the Lagrangian we assume a general expression of the form
\begin{equation}\label{General_L_continuum} 
L( \boldsymbol{\varphi }  , \dot{ \boldsymbol{\varphi }  })= \int_ \mathcal{B} \left[ \frac{1}{2} \varrho _0 | \dot{\boldsymbol{\varphi } }|- \varrho_0  \,\mathcal{E} ( \mathbb{F}  , \varrho_0 , S_0, G_0) \right] {\rm d}^3 \boldsymbol{X} ,
\end{equation}
where $\mathbb{F} (t, \boldsymbol{X}) =\nabla \boldsymbol{\varphi } (t, \boldsymbol{X})$ is the deformation gradient, $\varrho_0 ( \boldsymbol{X})$ and $S_0( \boldsymbol{X})$ are the mass and entropy density in the Lagrangian description, and $G_0( \boldsymbol{X})$ is a given co-metric on $ \mathcal{B} $ taken here to be the canonical one, i.e. $G_0^{AB}= \delta ^{AB}$. Here, the dot now denotes the partial derivative with respect to time $t$. In the second term, the specific energy $\mathcal{E}$ in material description depends on the deformation gradient $\mathbb{F}$ of the current configuration and also parametrically on $ \varrho _0,S_0, G_0$. Making this dependence explicit is crucial to discuss the notion of covariance in continuum mechanics, see \cite{marsden1994mathematical,GBMaRa12}. The expression $ \mathcal{E} $ is assumed to be invariant under the action of the group $ \operatorname{Diff}( \mathcal{B} )$ on $ \boldsymbol{\varphi }  $, $ \varrho _0$, $S_0$, $ G_0$ given by
\begin{equation}\label{right_action} 
\boldsymbol{\varphi }  \mapsto \boldsymbol{\varphi } \circ \boldsymbol{\phi}  , \qquad \varrho_0 \mapsto (\varrho_0  \circ \boldsymbol{\phi} ) J \boldsymbol{\phi} , \qquad S_0  \mapsto (S_0  \circ \boldsymbol{\phi} ) J \boldsymbol{\phi}, \qquad G_0 \mapsto \boldsymbol{\phi} _* G_0,
\end{equation}
for all $ \boldsymbol{\phi} \in \operatorname{Diff}( \mathcal{B})$, where $J \boldsymbol{\phi}  =| \operatorname{det}( \nabla \boldsymbol{\phi} )| $ is the Jacobian of $ \boldsymbol{\phi}$. In \eqref{right_action}, $ \boldsymbol{\phi} _*G_0$ denotes the push-forward  of the contravariant symmetric tensor field $G_0$ given in coordinates
\[
(\boldsymbol{\phi} _* G_0)^{AB}( \boldsymbol{X}) = G_0^{\,CD} (\boldsymbol{\phi} ^{-1}( \boldsymbol{X})) \frac{\partial \boldsymbol{\phi}^A}{\partial \boldsymbol{X} ^C} (\boldsymbol{\phi} ^{-1}( \boldsymbol{X}))\frac{\partial \boldsymbol{\phi}^B}{\partial \boldsymbol{X} ^D} (\boldsymbol{\phi} ^{-1}( \boldsymbol{X}))
\]

As a consequence of this invariance, one easily checks that the Lagrangian can be written in terms of Eulerian variables as
\begin{equation}\label{Eulerian_Lagrangian}
L( \boldsymbol{\varphi }  , \dot{ \boldsymbol{\varphi }  })= \int_ { \boldsymbol{\varphi }  ( \mathcal{B} )} \left[ \frac{1}{2} \rho  | \boldsymbol{u}  | ^2 - \rho\,  e ( \rho  , s, b) \right]  {\rm d}^3 \boldsymbol{x}  =:\ell( \boldsymbol{u}  , \rho  , s, b) ,
\end{equation} 
where $ \boldsymbol{u}$ is the Eulerian velocity, $ \rho,s  $ the Eulerian mass and entropy densities, and $b$ the Finger deformation tensor, defined by
\begin{align} 
\bu(t, \boldsymbol{x}) &= \dot{ \boldsymbol{\varphi }  } \big(t,  \boldsymbol{\varphi }  ^{-1}(t, \boldsymbol{x})\big)\label{Eulerian_velocity}\\
\rho (t, \boldsymbol{x}) &= \varrho _0 \big( \boldsymbol{\varphi }  ^{-1} (t, \boldsymbol{x})\big) J \boldsymbol{\varphi }  ^{-1}(t, \boldsymbol{x}) \label{Eulerian_mass}\\
s(t, \boldsymbol{x})&  = S _0 \big( \boldsymbol{\varphi }  ^{-1} (t, \boldsymbol{x})\big)J \boldsymbol{\varphi }  ^{-1}(t, \boldsymbol{x})\label{Eulerian_entropy}\\
b (t, \boldsymbol{x})&= \boldsymbol{\varphi }  _* G_0(t, \boldsymbol{x})\label{Eulerian_Finger}.
\end{align} 
When $G_0^{AB}= \delta ^{AB}$, the Finger deformation tensor has the standard expression
\begin{equation}\label{b_def} 
b(t, \boldsymbol{x})= \mathbb{F} (t, \boldsymbol{X})\mathbb{F} (t, \boldsymbol{X})^\mathsf{T}, \qquad \text{i.e.,} \qquad b^{ij}(t, \boldsymbol{x})=  \frac{\partial \boldsymbol{\varphi }  ^i}{\partial X^A}(t, \boldsymbol{X})\frac{\partial \boldsymbol{\varphi }  ^j}{\partial X^A}(t, \boldsymbol{X}),
\end{equation} 
where $ \boldsymbol{x}= \boldsymbol{\varphi }  (t, \boldsymbol{X})$.
In \eqref{Eulerian_Lagrangian} the function $e=e( \rho  , s, b)$ is the specific internal energy of the continuum in Eulerian description.

\subsubsection{Variational principles in absence of  reversible processes} In the material description and in absence of irreversible processes, the equations of motion are the Euler-Lagrange equations arising from the Hamilton principle
\begin{equation}\label{HP} 
\delta \int_0^T L( \boldsymbol{\varphi }  , \dot{\boldsymbol{\varphi } } ){\rm dt}=0
\end{equation}
with respect to arbitrary variations $ \delta \boldsymbol{\varphi }  $ of the embedding with $ \delta \boldsymbol{\varphi }  _{t=0,T}=0$.

In the Eulerian description, using \eqref{Eulerian_Lagrangian} and \eqref{Eulerian_velocity}--\eqref{Eulerian_Finger}, Hamilton's principle induces the following variational principle
\begin{equation}\label{EP} 
\delta \int_0^T\ell( \boldsymbol{u}, \rho  , s, b) {\rm d}t=0 
\end{equation}
for constrained variations
\begin{equation}\label{Eulerian_variations} 
\begin{aligned} 
\delta \boldsymbol{u}&= \partial _t \boldsymbol{\eta} + \boldsymbol{u} \cdot \nabla \boldsymbol{\eta} - \boldsymbol{\eta} \cdot \nabla \boldsymbol{u} \\
\delta   \rho  &= - \operatorname{div}( \rho  \boldsymbol{\eta} )\\
\delta   s  &= - \operatorname{div}( s  \boldsymbol{\eta} )\\
\delta b &= - \pounds _ {\boldsymbol{\eta}}b\, ,
\end{aligned}
\end{equation}  
where $ \boldsymbol{\eta} $ is an arbitrary time dependent vector field vanishing at $t=0,T$, see \cite{GBMaRa12}.
The expressions of the constrained variations \eqref{Eulerian_variations} are found by computing the variations of the variables in \eqref{Eulerian_velocity}--\eqref{Eulerian_Finger} induced by the variations $ \delta \boldsymbol{\varphi }  $ and defining $ \boldsymbol{\eta}= \delta \boldsymbol{\varphi }  \circ \boldsymbol{\varphi }  ^{-1} $. 
The last expression in \eqref{Eulerian_variations} is the Lie derivative of the symmetric two-contravariant tensor $b$ given in coordinates by
\begin{equation}\label{Lie_der_b}
(\pounds_{\boldsymbol{\eta} }b)^{ij}= \frac{\partial b^{ij}}{\partial x^k} \eta ^k - b^{kj}\frac{\partial \eta ^i }{\partial x^k} - b^{ik}\frac{\partial \eta ^j }{\partial x^k}\,. 
\end{equation}
A direct application of the variational principle \eqref{EP}--\eqref{Eulerian_variations} yields the equations of motion in Eulerian coordinates as
\begin{equation}\label{reduced_EL} 
\partial_t \frac{\delta\ell}{\delta \boldsymbol{u} }+\pounds_{\boldsymbol{u}}\frac{\delta\ell}{\delta\boldsymbol{u}} =\rho\nabla \frac{\delta\ell}{\delta\rho}  +s\nabla \frac{\delta\ell}{\delta s}- \frac{\delta\ell}{\delta b}:\nabla b- 2\operatorname{div} \left( \frac{\delta\ell}{\delta b}\cdot b \right),
\end{equation} 
see \cite{GBMaRa12}. The equations for $ \rho  , s, b$ follow from \eqref{Eulerian_mass}--\eqref{Eulerian_Finger} as
\[
\partial _t \rho  + \operatorname{div}( \rho  \boldsymbol{u}  )=0, \qquad \partial _t s  + \operatorname{div}( s  \boldsymbol{u}  )=0, \qquad \partial _t b + \pounds _{ \boldsymbol{u}} b=0.
\]
By using the expression of the Lagrangian given in \eqref{Eulerian_Lagrangian}, the equations of motion \eqref{reduced_EL} take the standard form
\[
\rho  (\partial _t \boldsymbol{u} + \boldsymbol{u} \cdot \nabla \boldsymbol{u}) = - \nabla p + \operatorname{div} \boldsymbol{\sigma}_{\rm el}, \qquad   p = \rho  ^2  \frac{\partial e}{\partial \rho  } , \qquad \boldsymbol{\sigma}_{\rm el} = 2 \rho  \frac{\partial e}{\partial b} \cdot b,
\] 
where $p$ is the pressure and $ \boldsymbol{\sigma} _{\rm el}$ is the elastic stress.
The boundary conditions will be discussed below in the case of porous media.

\subsubsection{Extension of this geometric variational setting} In this paper we shall use two types of extensions of the geometric variational formulation summarized above. First, we shall use the extension of \eqref{EP}--\eqref{Eulerian_variations} to porous media by following the approach of \cite{FaFGBPu2020}, then we shall further develop an extension to include irreversible processes, based on the variational formulation for nonequilibrium thermodynamics developed in \cite{gay2017lagrangian,gay2017lagrangian2}.

\color{black} 

\subsection{Definition of variables for porous media}

We present here the variables needed for the description of a porous media with internal entropies, by extending the approach described above.

\subsubsection{Configuration of the elastic body and the fluid} The motion of the elastic body (indexed by $s$) and the fluid (indexed by $f$) is defined by two time dependent maps $ \boldsymbol{\varphi }_s (t,\_\,): \mathcal{B} _s \rightarrow \mathbb{R} ^3  $ and $\boldsymbol{\varphi}_f (t,\_\,): \mathcal{B} _f \rightarrow \mathbb{R} ^3$ with variables denoted as 
\[
\boldsymbol{x} = \boldsymbol{\varphi }_s  (t,\bX) \quad\text{and}\quad \boldsymbol{x} = \boldsymbol{\varphi }_f (t,\bY)\,.
\]
Here $ \mathcal{B} _s$ and $ \mathcal{B} _f$ denote the reference configurations containing the elastic and fluid labels $\bX$ and $\bY$.
We assume that there is no fusion of either fluid or elastic body particles, so the map $ \boldsymbol{\varphi }_s$  and $ \boldsymbol{\varphi}_f $ are embeddings for all times $t$. 
We also assume that the fluid cannot escape the porous medium
or create voids, so at all times $t$, the domains occupied in space by the fluid  ${\mathcal B}_{t,f}= \boldsymbol{\varphi}_f (t,\mathcal{B})$ and the elastic body ${\mathcal B}_{t,s}= \boldsymbol{\varphi }_s (t,\mathcal{B})$ coincide: 
${\mathcal B}_{t,f}={\mathcal B}_{t,s}={\mathcal B}_{t}$. Finally, we shall assume for simplicity that the domain ${\mathcal B}_t$ does not change with time, and will simply call it ${\mathcal B}$, hence both $ \boldsymbol{\varphi} _f: \mathcal{B} _f \rightarrow \mathcal{B} $ and $ \boldsymbol{\varphi }_s : \mathcal{B} _s \rightarrow \mathcal{B}$ are diffeomorphisms for all time $t$. An extension to the case of the fluid escaping the boundary is possible, although it will require appropriate modifications in the variational principle and we shall consider it in a future work

\subsubsection{Velocities of the elastic body and the fluid} 
The fluid velocity $\bu_f$ and elastic solid velocity $\bu_s$, measured relative to the fixed coordinate system, \emph{i.e.}, in the Eulerian representation, are given as in the case of a single continuum in \eqref{Eulerian_velocity} by 
\begin{equation} 
\bu_f(t,\boldsymbol{x} )=\partial_t  \boldsymbol{\varphi} _f  \big(t, \bvarphi_f^{-1}(t,\boldsymbol{x})\big) \, , \quad  \bu_s(t,\boldsymbol{x})=\partial_t \boldsymbol{\varphi} _s \big(t, \boldsymbol{\varphi} _s^{-1}(t,\boldsymbol{x})\big) \, ,
\label{vel_def} 
\end{equation} 
for all $\boldsymbol{x} \in \mathcal{B}$.  Note that since $ \boldsymbol{\varphi} _f$ and $ \boldsymbol{\varphi} _s$ restrict to the boundaries, the vector fields $\bu_f$ and $\bu_s$ are tangent to the boundary, \emph{i.e.},
\begin{equation} \label{free_slip} 
\bu_f\cdot \boldsymbol{n} =0\, , \quad  \bu_s\cdot \boldsymbol{n} =0 \, ,
\end{equation}
where $ \boldsymbol{n}$ is the unit normal vector field to the boundary.  One can alternatively impose that $\bvarphi_f$ and $ \boldsymbol{\varphi }_s $ (or only $\boldsymbol{\varphi }_s $) are prescribed on the boundary. In this case, one gets no-slip boundary conditions
\begin{equation} \label{no_slip} 
\bu_f|_{ \partial \mathcal{B} }=0\, , \quad  \bu_s|_{ \partial \mathcal{B} }=0 \, , \quad \text{(or only $ \bu_s|_{ \partial \mathcal{B} }=0$)}.
\end{equation}

\subsubsection{Observed and actual densities} 
It is important to make a distinction between the observed density of the fluid or solid in a given volume, and the actual density of fluid filing the pores or elastic material comprising the matrix. The observed, or Eulerian density of fluid, is defined as the coefficient of proportionality between the mass of fluid contained in the given Eulerian volume $\mbox{d}^3 \boldsymbol{x}$, centered at the spatial point $ \boldsymbol{x}$, and the mass contained in that volume, and similarly for the solid: 
\begin{equation} 
\mbox{d} m_f (\boldsymbol{x},t) = \rho_f(\boldsymbol{x},t) \mbox{d}^3\boldsymbol{x}, 
\quad 
\mbox{d} m_s (\boldsymbol{x},t) = \rho_s(\boldsymbol{x},t) \mbox{d} ^3\boldsymbol{x} \, . 
\label{solid_def} 
\end{equation}
The actual density of the fluid and solid is the density of the material filling the pores (for example, density of gas in the pores for fluid) or, correspondingly, density of the elastic material comprising the matrix (\emph{e.g.}, rubber). The actual densities will be denoted with bars $\bar \rho_f$, $\bar \rho_s$. If $\phi(\boldsymbol{x},t)$ is the volume fraction of the pores, and we assume that the fluid fills the pores completely, the actual and Eulerian densities are related by 
\begin{equation} 
\rho_f = \phi \bar \rho_f \, , \quad \rho_s = (1-\phi) \bar \rho_s \, . 
\label{bar_no_bar} 
\end{equation} 

\subsubsection{Conservation law for the fluid and solid} 
Let us  look at the mass of fluid $\rho_f(t,\boldsymbol{x} ) \mbox{d}^3 \boldsymbol{x}$ , where $\rho_f$ is the observed (Eulerian) density of fluid. We shall consider both the actual density of fluid in pores $\bar \rho_f$ and the Eulerian density of fluid $\rho_f$, which are connected through $\rho_f=\phi \bar \rho_f$. The fluid must fill all the available volume completely, and it must have come from the material point  $\bY=\bvarphi^{-1}(t, \boldsymbol{x} )$. If the initial volume fraction at that point was $\rho_f^0(\bY) \mbox{d}^3 \bY$, then at a point $t$ in time we have 
\begin{equation} 
\rho_f(t,\boldsymbol{x}) = \rho_f^0 \big(\bvarphi_f^{-1}(t,\boldsymbol{x})\big)J\bvarphi_f^{-1}(t,\boldsymbol{x})  \, , \quad 
J\bvarphi_f^{-1} := |{\rm det} \big(\nabla \bvarphi_f^{-1}\big) |\, . 
\label{cons_law_fluid} 
\end{equation} 
Differentiating \eqref{cons_law_fluid}, we obtain the conservation law for $\rho_f(t,\boldsymbol{x})$ written as 
\begin{equation} 
\partial_t \rho_f + \operatorname{div}  ( \rho_f \boldsymbol{u} _f ) =0 \, . 
\label{rho_f_cons} 
\end{equation} 
The mass density of the elastic body, denoted $\rho_s$, satisfies an equation analogous to \eqref{cons_law_fluid}, namely,
\begin{equation} 
\rho_s(t,\boldsymbol{x}) = \rho_s^0\big( \boldsymbol{\varphi} _s ^{-1}(t,\boldsymbol{x})\big)J \boldsymbol{\varphi} _s ^{-1}(t,\boldsymbol{x})  \, ,
\label{cons_law_elastic} 
\end{equation} 
where $\rho_s^0(\bX)$ is the mass density in the reference configuration. The corresponding differentiated form of the conservation law is
\begin{equation} 
\partial_t \rho_s + \operatorname{div}  ( \rho_s \boldsymbol{u} _s ) =0 \, . 
\label{rho_s_cons} 
\end{equation}

\subsubsection{On microscopic variables, pore size and free volume}
Out of many microscopic variables presented in the porous matrix, the geometric shape of pores, and their connectivity are most important for computing the volume occupied by the fluid. For example, in  \cite{FaFGBPu2020}, as well as several papers before us \cite{placidi2008variational,serpieri2016variational,serpieri2017variational} and others, the internal 'microscopic' volume of the pores is chosen as an important variable affecting the potential energy of the solid. 
This choice is true for the case when the pores' geometry will be roughly similar throughout the material. The model will need to be corrected when there is a drastic change of pores' geometry (e.g. from roughly spherical to elliptical pores, for merging of pores \emph{etc}). The articles cited above consider that the locally averaged internal volume of the pores is represented by the local variable $v(t, \boldsymbol{x})$ in the Eulerian description, or its corresponding Lagrangian counterpart $\mathcal{V}(t,\bX) = v(t, \boldsymbol{\psi} (t,\bX))$.  In \cite{FaFGBPu2020}, the elastic energy of the solid  depends on the Finger deformation tensor $b$ and the infinitesimal pore volume $v$. Physically, this assumption is equivalent to stating that the internal volume variable $v$ will encompass all the effects of microscopic deformations on the elastic energy. 

The pore volume fraction $ \phi (t,\boldsymbol{x} )$ we have used is connected with the local concentration of pores $c(t,\boldsymbol{x})$ and the infinitesimal pore volume $v(t,\boldsymbol{x})$ as: 
\begin{equation} 
\phi(t, \boldsymbol{x} ) = c(b(t,\boldsymbol{x})) v(t,\boldsymbol{x}) \, . 
\label{g_c_v_constraint} 
\end{equation} 
If, for example, the pores are ``frozen'' in the material, they simply move as the material moves. Then, the change of the local concentrations of pores $c(t,\boldsymbol{x} )$ due to deformations is given by
\begin{equation}\label{change_c}
c\big(t, \boldsymbol{\varphi} _s (t,\bX)\big) J \boldsymbol{\varphi} _s(t,\bX) = c_0(\bX)\,, \quad J \boldsymbol{\varphi} _s = |{\rm det} ( \nabla \boldsymbol{\varphi} _s  )| =  | {\rm det}(\mathbb{F})|\,,
\end{equation}
where $c_0(\bX)$ is the initial concentration of pores in the Lagrangian point $\bX$. Using the definition \eqref{b_def} of the Finger tensor $b$ gives ${\rm det} \,b(t,\boldsymbol{x} ) = |{\rm det} \,\mathbb{F}(t,\bX)|^2$, hence we can rewrite the previous relation as
\[
c(t,\boldsymbol{x})\sqrt{{\rm det} \,b(t,\boldsymbol{x}) } = c_0(\bX)\,.
\]
In the case of an initially uniform porous media, \emph{i.e.}, $c_0= const$, this formula shows that the concentration $c(t,\boldsymbol{x})$ is a function of the value $b(t,\boldsymbol{x})$ of the Finger deformation tensor
\begin{equation}
c(b)=\frac{c_0}{\sqrt{{\rm det}b}} \, .
\label{c_b_particular_neq} 
\end{equation}
Note that from \eqref{change_c}, the concentration of pores satisfies
\begin{equation} 
\label{c_evolution} 
\partial_t c+ \operatorname{div}(c\bu_s)=0\,.
\end{equation}

We will use the variable $\phi$ in our description, rather than the description of pore concentration $c$ and pore's volume $v$. An equivalent theory may be constructed in terms of $v$, using the relations $\phi=\rho_f/\bar \rho_f$ and \eqref{g_c_v_constraint}, as well as the evolution of concentration using \eqref{c_evolution}, connecting to the theory developed in \cite{FaFGBPu2020,FaFGBPu2020_2}.

\subsubsection{Fluid incompressibility} 
If the fluid is incompressible, then $\bar\rho_f$ is advected as a scalar, $\partial _t \bar\rho_f+ \boldsymbol{u}_f \cdot \nabla \bar \rho  _f=0$. Therefore, equation \eqref{rho_f_cons} becomes 
\begin{equation} 
\partial_t \phi + \operatorname{div}  ( \phi\, \boldsymbol{u} _f ) =0 \, . 
\label{rho_f_cons_incompressible} 
\end{equation} 
Note that the incompressibility condition of the fluid \eqref{rho_f_cons_incompressible} \emph{does not} mean that $\operatorname{div}\bu_f=0$. That statement is only true for the case where no elastic matrix is present, \emph{i.e.}, for pure fluid. In the porous media case, a given spatial volume contains both fluid and elastic parts. The conservation of volume available to the fluid is thus given by \eqref{rho_f_cons_incompressible}.

\subsubsection{Fluid and solid incompressibility} 
We shall also note that for biological applications, such as sponges, the matrix is composed from water-filled cells so the matrix is incompressible, while still being elastic. If we put both $\partial _t \bar\rho_f+ \boldsymbol{u}_f \cdot \nabla \bar \rho  _f=0$ in \eqref{rho_f_cons} 
and $\partial _t \bar\rho_s+ \boldsymbol{u}_s \cdot \nabla \bar \rho  _s=0$ in \eqref{rho_s_cons}, we obtain 
\begin{equation}
 \left\{ 
 \begin{aligned} 
 & \partial_t \phi + \operatorname{div}  ( \phi\, \boldsymbol{u} _f ) 
 \\
 & \partial_t (1-\phi) + \operatorname{div}  ( (1-\phi) \,\boldsymbol{u} _f ) 
 \end{aligned} 
 \right. 
 \quad \Longrightarrow \quad 
 \operatorname{div} \left(\phi\, \boldsymbol{u} _f +  (1-\phi) \,\boldsymbol{u} _s \right) =0 \,.
 \end{equation} 
We refer the reader to \cite{FaFGBPu2020_2} for details.

\subsubsection{Entropy considerations} 
In addition to the previous variables, we also introduce the thermodynamic variables $s_f$ and $s_s$ which are the entropy densities of the fluid and the solid parts, respectively. We shall assume \emph{in this section only}  that there is no heat exchange between the solid and the fluid. Of course, that assumption is unphysical since the fluid and solid are intertwined on the microscopic level. The point of these additional variables is simply to state the definition and compute the variations satisfied by the entropies. The irreversible exchange of heat between the fluid and the solid media will be considered in Section~\ref{sec:thermodynamics}.  
In the absence of heat exchange, the entropy density (the amount of entropy in a given volume) evolves similarly to the mass density when advected by the corresponding media. The evolution equations for the entropy in the fluid and the solid are thus given by: 
\begin{equation} 
\partial_t s_f + \operatorname{div}(s_f  \boldsymbol{u} _f ) =0 \, , \qquad 
\partial_t s_s + \operatorname{div} ( s_s \boldsymbol{u} _s )=0\, .
\label{s_fs_evolution} 
\end{equation} 
We remind the reader that the ratios $\eta_{s,f}=s_{s,f}/\rho_{s,f}$ are called specific entropy. From \eqref{s_fs_evolution} and the mass conservation equations \eqref{rho_f_cons} and \eqref{rho_s_cons} it can be directly seen that the evolution equations for $\eta_{s,f}$ are:
\begin{equation} 
\partial _t \eta   _f + \boldsymbol{u} _f \cdot \nabla\eta  _f =0 , \qquad 
\partial _t \eta   _s + \boldsymbol{u}  _s \cdot \nabla\eta  _s =0 \,,
\label{eta_evolution} 
\end{equation} 
In particular, if $\eta  _f (t=0)=c$ is constant in space, then $\eta  _f (t)= c$, for all $t$.

\subsection{The Lagrangian function and the variational principle in spatial variables}\label{subsec_Lagr}

\subsubsection{Lagrangian} For classical elastic bodies, the potential energy in the spatial description depends on the Finger deformation tensor $b$, \emph{i.e.}, $V=V(\rho_s, b)$. In the porous media case, we consider the potential energy to depend on $b$ and the typical pore size $v$, and we write $V=V(\rho_s,b,v)$. To make our description more consistent, we express $\phi = 1-\rho_s/\bar \rho_s$. Then, using the concentration of pores $c=c(b)$ given by \eqref{c_b_particular_neq} and $v=\phi/c$, we can rewrite the potential energy of the solid as $V=V(\rho_s,\phi, b)$. We shall also include the entropies of the fluid and solid $s_f$, $s_s$ in the Lagrangian through the thermodynamics internal energies $e_{f}(\bar \rho_{f},s_f/ \rho  _f )$ and 
$e_{s}(\bar \rho_{s},s_s/ \rho   _s )$. Then, 
the Lagrangian of the porous medium is the sum of the kinetic energies of the fluid and elastic body minus the potential energy of the elastic deformations, and including the specific heat energy  for each material which is a new term compared to previous theory: 
\begin{equation}\label{Lagr_def}
\begin{aligned} 
&\ell(\bu_f,\bu_s,  \rho_f,  \rho_s, s_f, s_s, b,\phi,v)\\
&= \int_{\mathcal{B}}
 \Big[\frac{1}{2}  \rho_f|\bu_f|^2 - \rho_f e_f(\bar \rho_f,s_f/ \rho  _f) + \frac{1}{2}\rho_s  |\bu_s|^2 - \frac{1}{2} \rho_s e_s(\bar \rho_s,s_s/ \rho  _s, b) \Big] \, {\rm d}^3 \boldsymbol{x}  \,. 
\end{aligned}
\end{equation} 
Here, $\bar \rho_f$ and $\bar \rho_s$ is the density of actual fluid inside the pores and the material composing the elastic matrix as defined by \eqref{bar_no_bar}. 
We have separated these quantities in \eqref{Lagr_def}, contrary to our description in \cite{FaFGBPu2020} where we used the observed quantities $\rho_f=\phi \bar \rho_f$ and $\rho_s=(1-\phi) \bar \rho_s$. The description in terms of actual densities in the internal energies is more convenient for thermodynamic considerations, since $\bar \rho_f$ and $\bar \rho_s$ explicitly depend on entropy through the equations of state for a given material. We have also combined the thermal and elastic energy for the solid into a single internal energy function $e_s(\bar \rho_s,s_s/ \rho  _s, b)$ for convenience.

\begin{remark}
{\rm \textbf{(On the representation of soild's energy as a sum of elastic and thermal energies)} 
The reader may have noticed that when we have combined an elastic and thermal energy of the solid in \eqref{Lagr_def} in one term $\rho_s e_s(\bar \rho_s,s_s/ \rho  _s, b)$, we have implicitly assumed that the elastic energy $V(\rho,\phi,b)$ depends on $\rho_s$ and $\phi$ through the actual density of the fluid $\bar \rho_s=\rho_s/(1-\phi)$, and not on $\rho_s$ and $\phi$ individually. We will use this assumption throughout the paper, as it makes sense from the material science point of view. Mathematically, more general expressions for potential energy can be considered without difficulty leading to slightly different final equations of motions. However, for the sake of compactness, we will focus only on the case in \eqref{Lagr_def} that we consider physically relevant. 
}
\end{remark}

Note that the expression \eqref{Lagr_def} explicitly separates the contribution from the fluid and the elastic body in simple physically understandable terms. We will illustrate how to derive the equations of motion in case of fluid incompressibility for reference, since it is one of the most challenging cases for analysis. The incompressibility of fluid and/or solid without thermodynamics was in fact considered in \cite{FaFGBPu2020_2} and we refer the reader to that paper for complete analysis.

\noindent\begin{remark}
{\rm \textbf{(On the functional form of internal energies $e_{s,f}$ in \eqref{Lagr_def})} 
We now comment about the chosen nature of the dependence of the internal energies $e_f$ and $e_s$ on the material and entropy densities. From \eqref{Lagr_def}, we see that these energies depend on the reduced density $\bar \rho_{f,s}$ and not on the observed densities $\rho_{f,s}$. This is because we are considering the internal energy of each material separately. For example, for the fluid, the internal energy of the fluid itself inside the pore depends on the fluid density \emph{inside the pore}, which is $\bar \rho_f$. Thus, we take the energy density function per unit mass as $e_f=e_f(\bar \rho_f,s_f/\rho_f)$ as in \eqref{Lagr_def}. If the fluid filled the whole infinitesimal volume, its energy density function would have been $\bar \rho_f e_f$; however, since the fluid fills only the fraction $\phi$ of the volume, the energy density is $ \phi \bar \rho_f e_f = \rho_f e_f $, as in  \eqref{Lagr_def}. Similar argument applies to the solid part of the internal energy. 
Incidentally, this physical choice of variables for $e_f$ also affords the most elegant mathematical exposition of equations, as we will see below. 
}
\end{remark}

\subsubsection{Variational principle}
In the case when both the fluid and solid are compressible, the variational principle states that the equations of motion, in the absence of friction and external forces,  are obtained by the critical action principle with the action
\begin{equation} 
S= \int_0^T \ell(\bu_f,\bu_s, \rho_s,\rho_f, b,\phi) \mbox{d} t \,. 
\label{action_p}
\end{equation} 
The equations of motion, in the absence of  are then obtained through the variational principle 
\begin{equation} 
\de S =0 \, , \quad S \mbox{ given by \eqref{action_p}\,. } 
\label{var_principle_0} 
\end{equation} 
The constrained variations of the Eulerian variables induced by the free variations $\delta \boldsymbol{\varphi} _f$ , $\delta \boldsymbol{\varphi} _s$ vanishing at $t=0,T$  are computed through the right-invariant quantities  $\boldeta_{f}$ and $\boldeta_{s}$ defined, respectively, as 
\begin{equation} 
\boldeta_{f}=\de \bvarphi_f \circ \bvarphi_f^{-1} \, , \quad 
\boldeta_{s}=\de \boldsymbol{\varphi} _s   \circ \boldsymbol{\varphi} _s  ^{-1} \, .
\label{eta_def} 
\end{equation} 
The variations of physical quantities are then expressed in terms of $\boldeta_f$, $\boldeta_s$ as 
\begin{equation} 
\begin{aligned}
\de \bu_f &= \partial_t \boldeta_{f} + \bu_f  \cdot \nabla\boldeta_f - 
\boldeta_f \cdot  \nabla \bu_f \\ 
\de \bu_s &= \partial_t \boldeta_{s} + \bu_s  \cdot \nabla\boldeta_s - 
\boldeta_s \cdot  \nabla\bu_s \\
\delta\rho_f&= - \operatorname{div}(\rho_f\boldeta_f) \\ 
\delta\rho_s&= - \operatorname{div}(\rho_s\boldeta_s) \\ 
\de s_f &  = - \operatorname{div}(s_f \boldeta_f) \\ 
\de s_s &  = - \operatorname{div}(s_s \boldeta_s) \\ 
\delta b &= - \pounds_{\boldeta_s}b \, ,
\label{u_rho_b_var}
\end{aligned} 
\end{equation} 
and variation $\de \phi$ is arbitrary. Note that in \cite{FaFGBPu2020,FaFGBPu2020_2} we have set $\de v$ to be arbitrary, which is an equivalent description.

In the case of the boundary conditions \eqref{free_slip} it follows from \eqref{eta_def} that $\boldeta_{f}$ and $\boldeta_{s}$ are  arbitrary time dependent vector fields vanishing at $t=0,T$ and tangent to the boundary $\partial\mathcal{B}$:
\begin{equation} 
\label{non-permeable} 
\boldeta_s\cdot \boldsymbol{n} =0\, , \quad  \boldeta_f\cdot \boldsymbol{n} =0 \, . 
\end{equation}
In the case of no-slip boundary conditions \eqref{no_slip}, we have
\begin{equation} 
\label{no_slip_var} 
\boldeta_f|_{ \partial \mathcal{B} }=0\, , \quad  \boldeta_s|_{ \partial \mathcal{B} }=0 \, , \quad \text{(or only $\boldeta_s|_{ \partial \mathcal{B} }=0$)}.
\end{equation}

\subsubsection{Incorporation of external and friction forces}
In the models of the media that do not include thermodynamics, frictions forces, or any other forces, acting on the fluid $ \boldsymbol{f} _f$ and the media $ \boldsymbol{f} _s$ can be incorporated into the variational formulation by using the Lagrange-d'Alembert principle for external forces.  This principle reads
\begin{equation} 
\de S+ \int_0^T\!\!\int_{{\cal B}}\left(  \boldsymbol{f} _f \cdot \boldeta_f + \boldsymbol{f} _s \cdot \boldeta_s \right)  \mbox{d}^3 \boldsymbol{x}  \, \mbox{d} t =0 \, , \quad  
\label{Crit_action} 
\end{equation} 
where $S$ is defined in \eqref{action_p} and the variations are given by \eqref{u_rho_b_var}. 
Such friction forces are usually postulated from general physical considerations. If these forces are due exclusively to friction, the forces acting on the fluid and media at any given point must be equal and opposite, \emph{i.e.} $ \boldsymbol{f} _f=- \boldsymbol{f} _s$, in the Eulerian treatment we consider here. For example, for porous media, it is common to posit the friction law
\begin{equation} 
\boldsymbol{f} _f = - \boldsymbol{f} _s= \mathbb{K} (\bu_s - \bu_f)\,,
\label{Darcy_law} 
\end{equation} 
with $\mathbb{K}$ being a positive definite matrix potentially dependent on material parameters and variables representing the media. In particular, the matrix $\mathbb{K}$ depends on the local porosity, composition of the porous media, deformation and other variables, which should presumable be obtained from theoretical observations, see \cite{costa2006permeability} for discussion. 
\rem{ 
The general functional form of dependence of $\mathbb{K}$ on the variables should be of the form $\mathbb{K}=\mathbb{K}(b,g)$. 
For example, when  deformations of porous media are neglected, \emph{i.e.}, assuming an isotropic and a non-moving porous matrix with $b={\rm Id}$, Kozeny-Carman equation is often used, which in our notation is  written in the form  $\mathbb{K} = \kappa g^3/(1-g)^2$, with $\kappa$ being a constant, see \cite{costa2006permeability} for discussion.
In general, the derivation of the dependence of tensor $\mathbb{K}$ on variables $g$ and $b$ from the first principles is difficult, and should presumably be obtained from experimental observations.
} 
In this approach, the energy dissipates as heat and is simply withdrawn from the system with no effect on the dynamics. We present this consideration for completeness here. The core part of the article starts at Section~\ref{sec:thermodynamics}, where we outline how to incorporate the inclusion of the heat exchange into the dynamics by using a variational formulation for thermodynamics developed in \cite{gay2017lagrangian,gay2017lagrangian2}. In that approach, there is no need to treat the friction force as an external force in the variational principle. The friction force is then included as an internal force in the variational principle contributing directly to thermal effects.

We now turn our attention to the derivation of equations of motion in all combinations of compressible/incompressible fluids. 

\subsection{The case when both the fluid and solid are compressible}\label{sec:PM_comp}

\subsubsection{General equations}

\rem{
Let us consider the Lagrangian, which is in its general form written as 
\begin{equation}\label{General_form} 
\ell( \boldsymbol{u} _f, \boldsymbol{u}_s,\rho_f,\rho_s,s _s , s_f , b,\phi)=  \int_ \mathcal{B} \mathcal{L}( \boldsymbol{u},\boldsymbol{u}_s,\rho_f,\rho_s,s _s , s_f , b,\phi){\rm d}^3 \boldsymbol{x} \, . 
\end{equation}
The specific expression of the physically relevant Lagrangian for application to porous media is 
\begin{equation}\label{standard_Lagrangian} 
\begin{aligned}
&\mathcal{L}( \boldsymbol{u}_f,\boldsymbol{u}_s,\rho_f,\rho_s,s _s , s_f , b,\phi)\\
&=\frac{1}{2}\rho_f|\boldsymbol{u}_f|^2+ \frac{1}{2}\rho_s|\boldsymbol{u}_s|^2- \rho_f  e_f\left(\frac{\rho_f}{\phi}, \frac{s_f}{  \rho  _f } \right) - \rho_s e_s\left(\frac{\rho_s}{1-\phi},\frac{s_s}{\rho  _s } ,b\right),
\end{aligned} 
\end{equation} 
where $ \rho  _f$ and $\rho  _s$, are the mass densities per unit total volume, $ \bar{ \rho  }_f= \frac{\rho_f}{\phi}$ and $\bar{ \rho  _s}=\frac{\rho_s}{1-\phi}$ are the mass density per unit volume of the constituent, and $ \eta _k = \frac{s_k}{ \rho  _k }$, $k=f,s$ are the specific entropies.

}

Given a Lagrangian $\ell$, the variational principle \eqref{Crit_action}, with arbitrary variations  $\delta \phi $ and other variations given by \eqref{u_rho_b_var}, yields 
\begin{equation}\label{General_equations_compressible}
\left\{
\begin{array}{l}
\vspace{0.2cm}\displaystyle\partial_t \frac{\delta\ell}{\delta \boldsymbol{u}_f}+\pounds_{\boldsymbol{u}_f}\frac{\delta\ell}{\delta \boldsymbol{u}_f} =\rho_f\nabla \frac{\delta\ell}{\delta\rho_f}+s_f\nabla \frac{\delta\ell}{\delta s_f}+ \boldsymbol{f} _f\\
\vspace{0.2cm}\displaystyle\partial_t \frac{\delta\ell}{\delta \boldsymbol{u}_s}+\pounds_{ \boldsymbol{u}_s}\frac{\delta\ell}{\delta \boldsymbol{u}_s} =\rho_s\nabla \frac{\delta\ell}{\delta\rho_s} +s_s\nabla \frac{\delta\ell}{\delta s_s}- \frac{\delta\ell}{\delta b}\!:\!\nabla b- 2\operatorname{div} \left( \frac{\delta\ell}{\delta b}\!\cdot\! b \right)+ \boldsymbol{f} _s \\
\vspace{0.2cm}\displaystyle\partial_t \rho_f+ \operatorname{div}(\rho_f \boldsymbol{u}_f)=0,\qquad \partial_t \rho_s+ \operatorname{div}(\rho_s \boldsymbol{u}_s)=0,\qquad\partial_t b+ \pounds_{\boldsymbol{u}_s}b=0\\
\vspace{0.2cm}\displaystyle\partial_t s_f+ \operatorname{div}(s_f \boldsymbol{u}_f)=0,\qquad \partial_t s_s+ \operatorname{div}(s_s \boldsymbol{u}_s)=0\\
\displaystyle\frac{\delta\ell}{\delta\phi}=0.
\end{array}
\right.
\end{equation}
By using the Lagrangian $\ell$ defined by \eqref{Lagr_def} above, we get the system of equations 
\begin{equation}\label{classic_porousmedia}
\left\{
\begin{array}{l}
\vspace{0.2cm}\displaystyle \rho_f (\partial_t \mathbf{u}_f+ \mathbf{u}_f\cdot \nabla \mathbf{u}_f) = - \phi \nabla p + \boldsymbol{f}_f \\
\vspace{0.2cm}\displaystyle\rho_s( \partial_t \mathbf{u}_s+ \mathbf{u}_s\cdot \nabla \mathbf{u}_s) = - (1-\phi) \nabla p +  \operatorname{div} \boldsymbol{\sigma} _{\rm el} + \boldsymbol{f} _s\\
\vspace{0.2cm}\displaystyle\partial_t \rho_f+ \operatorname{div}(\rho_f \mathbf{u}_f)=0\,,\qquad \partial_t \rho_s+ \operatorname{div}(\rho_s \mathbf{u}_s)=0\,,\qquad\partial_t b+ \pounds_{\mathbf{u}_s}b=0\\
\partial_t s_f + \operatorname{div} ( s_f \bu_f ) =0 \, , \qquad 
\partial_t s_s + \operatorname{div} ( s_s \bu_s  )=0\, ,
\end{array}
\right.
\end{equation}
where we have defined the pressure $p$ as the two equal terms coming from the conditions $\de \ell/\de \phi=0$: 
\begin{equation} 
p:=\displaystyle  \bar\rho_f^2\frac{\partial e_f}{\partial\bar\rho_f} =  \bar\rho_s^2\frac{\partial e_s}{\partial\bar\rho_s}\,.
\label{p_compressible} 
\end{equation} 
Hence $ \delta \ell/ \delta \phi =0$ imposes that both components have the same pressure. We have also defined the elastic stress as 
\begin{equation} 
\label{sigma_el_def}
\boldsymbol{\sigma} _{\rm el}:= 2 \rho_s \frac{\partial e_s}{\partial  b}\cdot b\,.
\end{equation} 

From condition \eqref{p_compressible} we get an algebraic expression for the volume fraction of the pores in terms of the mass and entropy densities and of the Finger deformation tensor as $ \phi = \phi( \rho  _f, s_f, \rho  _s, s_s, b)$, which is then inserted in the two momentum equations in \eqref{classic_porousmedia}, thereby giving a system of seven equations for the variables $ \boldsymbol{u}_f, \boldsymbol{u}_s, \rho  _f, s_f, \rho  _s, s_s, b$.

In the case of the boundary conditions \eqref{free_slip}, by using \eqref{non-permeable} the variational principle gives
\[
(\boldsymbol{\sigma} _{\rm el} \cdot \boldsymbol{n}  ) \cdot \boldsymbol{\eta} _s=0\,,
\]
for all $ \boldsymbol{\eta} _s $ parallel to the boundary, i.e, $\boldsymbol{\sigma} _{\rm el} \cdot \boldsymbol{n} = \lambda \boldsymbol{n}  $, for some $ \lambda$. When $ \boldsymbol{u}  _s |_{ \partial \mathcal{B} }=0$ no further condition arise from the variational principle, see \eqref{no_slip_var}.

Equations \eqref{classic_porousmedia} do provide a description of porous media with trivial thermodynamics effects, caused by the presence of an entropy which is simply advected. However, these equations involve no irreversible thermodynamics effects, such as heat exchange between fluid and solid, which is unrealistic. 
For example, it is hard to incorporate consistently how the thermal dissipation is affecting the entropy production during the complex motion we are considering here. A more complex and consistent theory will be derived in Section~\ref{sec:thermodynamics} below. In the meantime, we will explore some particular cases for completeness of exposition. In addition, the thermodynamics approach we develop in Section~\ref{sec:thermodynamics} allows to deduce additional information about the physically relevant structure of dissipative forces which is impossible to guess in this framework.

\rem{
By using the Lagrange-d'Alembert principle with the virtual force term
\[
\int_{ \partial \mathcal{B}} \boldsymbol{t}  \circ \boldsymbol{\psi}  \cdot \delta  \boldsymbol{\psi}  {\rm d}s= \int _ { \partial \mathcal{B}} \boldsymbol{t}  \cdot \boldsymbol{\eta} _s {\rm d}s 
\]
for some given traction $\boldsymbol{t} $, we get the boundary condition
\[
(\boldsymbol{\sigma} _{\rm el} \cdot \boldsymbol{n} - \boldsymbol{t}  ) \cdot \boldsymbol{\eta} _s=0,
\]
for all $ \boldsymbol{\eta} _s $ parallel to the boundary, i.e, $(\boldsymbol{\sigma} _{\rm el} \cdot \boldsymbol{n} - \boldsymbol{t} )= \lambda \boldsymbol{n}  $, for some $ \lambda$. This is consistent with the fact that only the component of the traction force parallel to the boundary counts.
}

\subsection{The case of an incompressible fluid}

\subsubsection{Background of the approach} 
\label{sec:incompr_fluid}

Condition \eqref{rho_f_cons_incompressible} or its integrated version $\phi= (\phi_0 \circ \boldsymbol{\varphi } _f ^{-1} )J \boldsymbol{\varphi} _f ^{-1}$ represents a scalar constraint for every point of an infinite-dimensional system. Formally, such constraint can be treated in terms of Lagrange multipliers. The application of the method of Lagrange multipliers for an infinite-dimensional system is quite challenging, see the recent review papers \cite{dell2018lagrange,bersani2019lagrange}. In terms of classical fluid flow, in the framework of Euler equations, the variational theory introducing incompressibility constraint has been developed by V. I. Arnold \cite{arnold1966geometrie} on diffeomorphism groups, with the Lagrange multiplier for incompressibility related to the pressure in the fluid. We will follow in the footsteps of Arnold's method and introduce a Lagrange multiplier for the incompressibility condition \eqref{rho_f_cons_incompressible}. By analogy with Arnold, we will also treat this Lagrange multiplier as related to pressure, as it has the same dimensions.  Since \eqref{rho_f_cons_incompressible} refers to the fluid content, the Lagrange multiplier relates to the pressure of the fluid. The equations of motion \eqref{General_equations_incompressible} and, consequently, \eqref{p_compressible_incompressible}, derived below, connecting pressure with the derivatives of the potential energy with respect to the pores' volume, will further justify this concept. Note that the Lagrange multiplier may be different from the actual physical pressure in the fluid depending on the implementation of the model.

From the Lagrangian \eqref{Lagr_def} and the constraint $\phi= (\phi_0 \circ \boldsymbol{\varphi } _f ^{-1} )J \boldsymbol{\varphi} _f ^{-1}$, we define the action functional in the Eulerian description as
\begin{equation} 
S= \int_0^T\bigg[ \ell(\bu_f,\bu_s, \rho_s,\rho_f, s_f, s_s, b,\phi)+  \int_{{\cal B}} \mu \left(  \phi- (\phi_0 \circ \boldsymbol{\varphi } _f ^{-1} )J \boldsymbol{\varphi} _f ^{-1} \right)  \mbox{d}^3 \boldsymbol{x} \bigg] \mbox{d} t \,. 
\label{action_p_new}
\end{equation} 
Note the difference with \eqref{action_p}--\eqref{var_principle_0}  for the fully compressible case.

\begin{remark}
{\rm 
Note that in the compressible case, we derive the pressure by taking derivatives of the energy of the fluid and solid, $e_f$ and $e_s$ respectively, with respect to $\bar \rho_s$ and $\bar \rho_f$. This is the approach to pressure also taken in \cite{ChMo2010,ChMo2014,vuong2015general}, only reformulated in the spatial variables. In contrast, in the incompressible case, we treat the pressure as a Lagrange multiplier for the incompressibility condition, which is an approach different from \cite{ChMo2010,ChMo2014,vuong2015general}}
\end{remark}

\subsubsection{General equations}

The equations of motion are obtained by computing the critical points condition
\begin{equation} 
\label{incompressible_fluid}
\begin{aligned} 
&\delta \int_0^T \bigg[\ell( \boldsymbol{u} _f,\boldsymbol{u}_s,\rho_f,\rho_s,s _f , s_s , b,\phi)+  \int_\mathcal{B} \mu  \Big( \phi - (\phi^0 \circ \boldsymbol{\varphi} _f ^{-1})J \boldsymbol{\varphi}_f^{-1}\Big){\rm d}^3 \boldsymbol{x} \bigg]{\rm d}t\\
&\hspace{6cm}+ \int_0^T\int_ \mathcal{B} \left(  \boldsymbol{f} _f \cdot \boldeta_f  +  \boldsymbol{f} _s \cdot   \boldeta_s \right) \mbox{d}^3\boldsymbol{x} \mbox{d} t =0 \,,
\end{aligned}
\end{equation}
with respect to free variations $ \delta \mu $ and $\delta \phi$ and with respect to the constrained variations \eqref{u_rho_b_var} of the Eulerian variables induced by free variations of the Lagrangian variables.
Using the variation
\begin{equation} 
\label{phi_identity} 
\delta \Big((\phi^0 \circ \boldsymbol{\varphi} _f ^{-1})J \boldsymbol{\varphi}_f^{-1}\Big)= - \operatorname{div}\Big((\phi^0 \circ \boldsymbol{\varphi} _f ^{-1} )J \boldsymbol{\varphi}_f^{-1} \boldsymbol{\eta}_f\Big) = - \operatorname{div} ( \phi\, \boldsymbol{\eta}_f),
\end{equation} 
with $ \boldsymbol{\eta} _f = \delta \boldsymbol{\varphi} _f \circ \boldsymbol{\varphi} _f ^{-1} $, we obtain the following equations for a general Lagrangian $\ell$: 
\begin{equation}\label{General_equations_incompressible} 
\left\{
\begin{array}{l}
\vspace{0.2cm}\displaystyle\partial_t \frac{\delta\ell}{\delta \boldsymbol{u}_f}+\pounds_{ \boldsymbol{u}_f}\frac{\delta\ell}{\delta \boldsymbol{u}_f} = \rho_f\nabla \frac{\delta\ell}{\delta\rho_f}  +s_f\nabla \frac{\delta\ell}{\delta s_f}- \phi \nabla \mu + \boldsymbol{f}_f \\
\vspace{0.2cm}\displaystyle\partial_t \frac{\delta\ell}{\delta \boldsymbol{u}_s}+\pounds_{ \boldsymbol{u}_s}\frac{\delta\ell}{\delta \boldsymbol{u}_s} =\rho_s\nabla \frac{\delta\ell}{\delta\rho_s} +s_s\nabla \frac{\delta\ell}{\delta s_s}- \frac{\delta\ell}{\delta b}:\nabla b- 2 \left( \operatorname{div} \frac{\delta\ell}{\delta b}\cdot b \right) + \boldsymbol{f}_s \\
\vspace{0.2cm}\displaystyle\partial_t \rho_f+ \operatorname{div}(\rho_f \boldsymbol{u}_f)=0\,,\qquad \partial_t \rho_s+ \operatorname{div}(\rho_s  \boldsymbol{u}_s)=0,\qquad\partial_t b+ \pounds_{ \boldsymbol{u}_s}b=0\\
\vspace{0.2cm}\partial_t s_f + \operatorname{div} ( s_f \boldsymbol{u}_f ) =0 \, , \qquad 
\partial_t s_s + \operatorname{div} ( s_s \boldsymbol{u}_s   )=0 \\
\displaystyle  \phi = (\phi^0 \circ \boldsymbol{\varphi} _f ^{-1})J \boldsymbol{\varphi}_f^{-1},\qquad \frac{\delta\ell}{\delta\phi}+  \mu =0\,.\\
\end{array}
\right.
\end{equation}
The constraint $\phi = (\phi^0 \circ \boldsymbol{\varphi} _f ^{-1})J \boldsymbol{\varphi}_f^{-1}$  means that
\begin{equation}\label{phi_equation_f}
\partial _t \phi + \operatorname{div}( \phi \boldsymbol{u}_f)=0  
\end{equation} 
and hence, from $ \bar \rho  _f= \rho  _f/ \phi $, the actual mass density is advected as a scalar as
\begin{equation}\label{rho_f_equation_inc} 
\partial _t \bar \rho  _f + \boldsymbol{u}_f \cdot \nabla \bar \rho  _f=0.
\end{equation} 
With the particular form of the physically relevant Lagrangian \eqref{Lagr_def}, the system of equations take \emph{exactly} the same form as equations \eqref{classic_porousmedia}  with the following two important changes induced by the Lagrange multiplier term in the variational principle:
\begin{itemize}
\item[(1)] there is the additional equation \eqref{phi_equation_f} governing the pore volume fraction;

\item[(2)] the pressure equation \eqref{p_compressible} is modified as  
\rem{ 
\begin{equation}\label{classic_porousmedia_incomp}
\left\{
\begin{array}{l}
\vspace{0.2cm}\displaystyle \rho_f (\partial_t \mathbf{u}_f+ \mathbf{u}_f\cdot \nabla \mathbf{u}_f) = - \phi \nabla p \\
\vspace{0.2cm}\displaystyle\rho_s( \partial_t \mathbf{u}_s+ \mathbf{u}_s\cdot \nabla \mathbf{u}_s) = - (1-\phi) \nabla p  +  \operatorname{div} \boldsymbol{\sigma} _{\rm el}\\
\vspace{0.2cm}\displaystyle\partial_t \rho_f+ \operatorname{div}(\rho_f \mathbf{u}_f)=0,\qquad \partial_t \rho_s+ \operatorname{div}(\rho_s \mathbf{u}_s)=0,\qquad\partial_t b+ \pounds_{\mathbf{u}_s}b=0\\
\vspace{0.2cm}\textcolor{red}{\displaystyle\partial_t s_f+
 \mathbf{u}_f \cdot \nabla s_f =0 , \qquad 
 \partial_t s_s+
 \mathbf{u}_s \cdot \nabla s_s =0 }
 \\
\displaystyle \partial_t \phi+ \operatorname{div}(\phi \mathbf{u}_f)=0,\qquad   \bar\rho_f^2\frac{\partial e_f}{\partial\bar\rho_f} + \mu = \bar\rho_s^2\frac{\partial e_s}{\partial\bar\rho_s}=:p
\end{array}
\right.
\end{equation} 
} 
\begin{equation} 
p:=\displaystyle  \bar\rho_f^2\frac{\partial e_f}{\partial\bar\rho_f}+ \mu =  \bar\rho_s^2\frac{\partial e_s}{\partial\bar\rho_s}\,.
\label{p_compressible_incompressible} 
\end{equation}
\end{itemize}
From the condition \eqref{p_compressible_incompressible} and using the expression of $e_f$ and $e_s$ we directly get the Lagrangian multiplier as a function $ \mu = \mu ( \rho  _f, s_f, \rho _s, s_s, b, \phi )$ that can be inserted in the two momentum equations. This gives a closed system of eight equations for the variables $ \boldsymbol{u}_f, \boldsymbol{u}_s, \rho  _f, s_f, \rho  _s, s_s,  b, \phi$.

From the two momentum equations and the equation \eqref{p_compressible_incompressible} it is also clear that the internal energy term of the fluid $ \rho  _f e_f( \bar \rho  _f, s_f/ \rho  
_f)$ in the Lagrangian only contributes to a redefinition of the Lagrange multiplier $ \mu $ and does not affect the equations which only involves $p$. The internal energy of the fluid, \emph{i.e.}, the term $\rho_f e_f$, can thus be discarded in the Lagrangian \eqref{Lagr_def} for the case of a porous media with incompressible fluid, and the fluid entropy equation can be dropped since it decouples from the other equations. The same properties are well known to happen for a single incompressible fluid, possibly with variable density. Note that when $e_f$ is dropped $ \mu $ is a function of $ \rho  _s, s_s, b, \phi $ only.

From \eqref{phi_equation_f} and the $ \rho  _f$ equation, the actual density $\bar \rho  _f$ of the fluid is advected as a scalar as we have seen in \eqref{rho_f_equation_inc}. 
Hence if $\bar \rho  _f( t=0)=c$ is constant in space, then $\bar \rho  _f(t)=c$ for all $t$. This corresponds to the case of a homogeneous incompressible fluid.

The boundary conditions and energy balances are derived similar to the calculations in \S\ref{sec:PM_comp}

\subsection{The case when both the fluid and solid are incompressible} \label{sec:double_incompr}

The case of having both the fluid and the solid matrix to be incompressible was considered in the context of biological systems in \cite{FaFGBPu2020_2}, without the thermodynamics consideration. 
In the case when thermodynamics is relevant, we take the Lagrange-d'Alembert action principle to be enforcing both the incompressibility of the fluid and the solid using the Lagrange multipliers $\mu_f$ and $\mu_s$ as  
\begin{equation} 
\begin{aligned} 
& \de  \int_0^T \bigg[ \ell( \boldsymbol{u}  _f , \boldsymbol{u}  _s , \rho  _f , \rho  _s , b,g, s_f,s_s)  + \int_ \mathcal{B}  \mu_f \left( \phi - (\phi^0 \circ \boldsymbol{\varphi} _f ^{-1})J \boldsymbol{\varphi} _f^{-1}\right) \mbox{d}^3 \boldsymbol{x} \\
& \hspace{3cm} + \int_ \mathcal{B} \mu_s \left( (1-\phi) - ((1-\phi^0) \circ \boldsymbol{\varphi} _s ^{-1})J \boldsymbol{\varphi} _s^{-1}\right)  {\rm d} ^3 \boldsymbol{x} \bigg]\mbox{d} t \\
& \hspace{3cm}+ \int_0^T\int_ \mathcal{B} \left(  \boldsymbol{f} _f \cdot \boldeta_f +  \boldsymbol{f} _s \cdot   \boldeta_s \right) \mbox{d}^3\boldsymbol{x} \mbox{d} t =0 \,, 
\end{aligned} 
\label{Crit_action_incompr_incompr} 
\end{equation} 
with arbitrary variations $ \delta \phi$,  $\delta \mu_f$, and $\delta \mu_s$, and the same Lagrangian as before. Using  \eqref{Crit_action_incompr_incompr}, for a general Lagrangian $\ell$, we obtain the following equations of motion: 
\begin{equation}\label{General_equations_incompressible} 
\left\{
\begin{array}{l}
\vspace{0.2cm}\displaystyle\partial_t \frac{\delta\ell}{\delta\mathbf{u}_f}+\pounds_{\mathbf{u}_f}\frac{\delta\ell}{\delta\mathbf{u}_f} = \rho_f\nabla \frac{\delta\ell}{\delta\rho_f}  +s_f\nabla \frac{\delta\ell}{\delta s_f}- \phi \nabla \mu_f + \boldsymbol{f} _f \\
\vspace{0.2cm}\displaystyle\partial_t \frac{\delta\ell}{\delta\mathbf{u}_s}+\pounds_{\mathbf{u}_s}\frac{\delta\ell}{\delta\mathbf{u}_s} =\rho_s\nabla \frac{\delta\ell}{\delta\rho_s} +s_s\nabla \frac{\delta\ell}{\delta s_s}-(1-\phi) \nabla \mu_s \\
\vspace{0.2cm}\displaystyle \hspace{7cm}- \frac{\delta\ell}{\delta b}:\nabla b- 2 \left( \operatorname{div} \frac{\delta\ell}{\delta b}\cdot b \right) + \boldsymbol{f} _s\\
\vspace{0.2cm}\displaystyle\partial_t \rho_f+ \operatorname{div}(\rho_f \mathbf{u}_f)=0,\qquad \partial_t \rho_s+ \operatorname{div}(\rho_s \mathbf{u}_s)=0\,,\qquad\partial_t b+ \pounds_{\mathbf{u}_s}b=0\\
\vspace{0.2cm}\partial_t s_f + \operatorname{div} ( s_f\bu_f) =0 \, , \qquad 
\partial_t s_s + \operatorname{div} ( s_s\bu_s  )=0\\
\vspace{0.2cm}\displaystyle  \phi = (\phi^0 \circ \boldsymbol{\varphi} _f ^{-1})J \boldsymbol{\varphi}_f^{-1}, \qquad 
 (1-\phi) = ((1-\phi^0) \circ \boldsymbol{\varphi} _s ^{-1})J \boldsymbol{\varphi} _s^{-1}\\
 \displaystyle \frac{\delta\ell}{\delta\phi}+  \mu_f - \mu_s  =0\,.\\
\end{array}
\right.
\end{equation} 
Using the expression for the Lagrangian given by  \eqref{Lagr_def}, we again get \emph{exactly} the system of equations \eqref{classic_porousmedia}, with the following important two changes induced by the two Lagrange  multiplier terms in  the  variational  principle:
\begin{itemize}
\item[(1)] there  are  the two additional equations governing the volume fractions
\begin{equation}\label{phi_equations_incom_incomp} 
\partial _t \phi + \operatorname{div}( \phi \, \boldsymbol{u}_f)=0, \qquad   \partial _t (1-\phi) + \operatorname{div}( (1-\phi )\, \boldsymbol{u}_s)=0\,,
\end{equation}
\item[(2)] the pressure equation now reads:
\rem{ 
\begin{equation}\label{classic_porousmedia_incomp_incompr}
\left\{
\begin{array}{l}
\vspace{0.2cm}\displaystyle \rho_f (\partial_t \mathbf{u}_f+ \mathbf{u}_f\cdot \nabla \mathbf{u}_f) = - \phi \nabla p + \mathbf{F}_f \\
\vspace{0.2cm}\displaystyle\rho_s( \partial_t \mathbf{u}_s+ \mathbf{u}_s\cdot \nabla \mathbf{u}_s) =- (1-\phi) \nabla p  +  \operatorname{div} \boldsymbol{\sigma} _{\rm el}+ \mathbf{F}_s\\
\vspace{0.2cm}\displaystyle\partial_t \rho_f+ \operatorname{div}(\rho_f \mathbf{u}_f)=0,\qquad \partial_t \rho_s+ \operatorname{div}(\rho_s \mathbf{u}_s)=0,\qquad\partial_t b+ \pounds_{\mathbf{u}_s}b=0\\
\vspace{0.2cm}\textcolor{red}{\displaystyle\partial_t s_f+
 \mathbf{u}_f \cdot \nabla s_f =0 , \qquad 
 \partial_t s_s+
 \mathbf{u}_s \cdot \nabla s_s =0 }\\
\vspace{0.2cm} \displaystyle \partial_t \phi+ \operatorname{div}(\phi \mathbf{u}_f)=0,\qquad  
 \partial_t (1-\phi)+ \operatorname{div}((1-\phi) \mathbf{u}_s)=0 
 \\
 \vspace{0.2cm}\displaystyle  \bar\rho_f^2\frac{\partial e_f}{\partial\bar\rho_f} + \mu_f = \bar\rho_s^2\frac{\partial e_s}{\partial\bar\rho_s}+\mu_s =:p
\end{array}
\right.
\end{equation} 
} 
\begin{equation} 
p:=\displaystyle  \bar\rho_f^2\frac{\partial e_f}{\partial\bar\rho_f}+ \mu_f =  \bar\rho_s^2\frac{\partial e_s}{\partial\bar\rho_s}+\mu_s\, .
\label{p_incompressible_incompressible} 
\end{equation} 
\end{itemize}
This pressure equation, obtained from the variations $\delta \phi$,  defines the effective pressure $p$ expressed in terms of two Lagrange multipliers $(\mu_f,\mu_s)$ enforcing the incompressibility of fluid and solid, respectively.
To solve the system, we note that the equations in \eqref{phi_equations_incom_incomp} are equivalently written as
\[
\partial _t \phi + \operatorname{div}( \phi \, \boldsymbol{u}_f)=0, \qquad   \operatorname{div}( \phi \boldsymbol{u}_f + ( 1- \phi ) \boldsymbol{u} _s)=0\,.
\]
By using the second equation above and the two momentum equations we get the Poisson pressure equation for $p$ in terms of the variables $ \boldsymbol{u}_f, \boldsymbol{u}_s, \rho  _f, s_f, \rho  _s, s_s, b, \phi $ whose dynamics are governed by the equations of system \eqref{classic_porousmedia} together with the first equation in \eqref{phi_equations_incom_incomp}. The Lagrange multiplier $ \mu _f$ and $ \mu _s$ are then found from \eqref{p_incompressible_incompressible}. As earlier, the fluid entropy equation decouples and the fluid internal energy term can be discarded in the Lagrangian, in which case the Lagrange multiplier $ \mu _f$ coincides with the actual pressure $p$. In addition, terms independent of $b$ (if any) in the solid internal energy term can also be discarded in the Lagrangian since they amount in a redefinition of $  \mu _s$ without changing the momentum equation.

\rem{ 
While it seems that there are two pressures and two Lagrange multipliers for the incompressibility conditions $\mu_s$ and $\mu_f$, the situation is actually much simpler. The key is to notice that we can define a unifying quantity $p$, with the physical meaning of pressure, according to  
\begin{equation} 
p:=\bar\rho_f^2\frac{\partial e_f}{\partial\bar\rho_f} + \mu_f = \bar\rho_s^2\frac{\partial e_s}{\partial\bar\rho_s}+\mu_s
\label{pressure_def} 
\end{equation} 
Using that definition of pressure, the first two dynamic equations for the evolution of fluid and solid in \eqref{classic_porousmedia_incomp_incompr} reduce to
\begin{equation}\label{classic_porousmedia_incomp2_final}
\left\{
\begin{array}{l}
\vspace{0.2cm}\displaystyle \rho_f (\partial_t \mathbf{u}_f+ \mathbf{u}_f\cdot \nabla \mathbf{u}_f) = - \phi \nabla  p + \mathbf{F}_f\\
\vspace{0.2cm}\displaystyle\rho_s( \partial_t \mathbf{u}_s+ \mathbf{u}_s\cdot \nabla \mathbf{u}_s) =- (1-\phi) \nabla p   +  \operatorname{div} \boldsymbol{\sigma} _{\rm el}
+ \mathbf{F}_s
\end{array}
\right.
\end{equation} 
} 

Now from \eqref{phi_equations_incom_incomp} both the actual densities $\bar \rho_f$ and $\bar \rho_s$ are advected by the corresponding velocities:  
\begin{equation} 
\partial_t \bar \rho_s + \boldsymbol{u} _s \cdot \nabla \bar \rho_s =0 \, , \qquad 
\partial_t \bar \rho_f + \boldsymbol{u} _f \cdot \nabla \bar \rho_f =0 \, . 
\label{rhofs_evolution} 
\end{equation} 
Thus, if $\rho_s(t=0)$ and $\rho_f(t=0)$ are constant in space, then they remain constant at all times. This corresponds to homogeneous fluid and solid and considerably simplifies the pressure Poisson equation. 
\color{black}

\rem{ 
The  two equations in the next to last line of \eqref{General_equations_incompressible}, upon addition, give the incompressibility constraint expressed for velocities: 
\begin{equation} 
 \operatorname{div} \left( \phi \mathbf{u}_f + (1-\phi) \mathbf{u}_s \right) =0 
\label{incompr_constr} 
\end{equation} 

The pressure $p$ appearing in the first two equations is computed from the elliptic equation deduced from taking the time derivative of the incompressibility constraint \eqref{incompr_constr}. The Lagrange multipliers $\mu_f$ and $\mu_s$ are then computed from the last equation. As earlier, the internal energy of the fluid can be neglected without changing the dynamics.

\todo{I suggest dropping above discussion of \eqref{incompr_constr} since we already have it in the introduction. } 
} 



\section{Porous media with thermodynamics}
\label{sec:thermodynamics} 

In this Section, we will develop a novel theory of poromechanics with thermodynamics exchange, based on the variational thermodynamics approach developed in \cite{gay2017lagrangian,gay2017lagrangian2,GBYo2019}. We start with a simple illustrative example of two pistons connected by a constraint, such as an inextensible rope. Simple as this problem is, it illustrates all essential features of the poromechanics problem, namely, two Lagrangian systems with thermodynamics connected through a holonomic constraint of incompressibility. This approach will allow us to compute a much more detailed expression for the friction force than was possible by guessing the friction terms $ \boldsymbol{f} _{f,s}$ in the Lagrange-d'Alembert principle \eqref{Crit_action}. Also, in addition to the friction forces, we can compute the general Navier-Stokes-like  friction stresses in the media. 

\subsection{A simple example of two connected pistons}\label{subsec_example_piston}

We present here the extension of the variational formulation \eqref{VaCond}--\eqref{VaCo} to systems experiencing internal heat conduction. A continuum analogue will be developed in \S\ref{sec:Thermo_comp} as a modeling tool for the thermodynamics of porous media.

\subsubsection{Thermodynamic displacement variables} An important variable for the extension of the variational formulation of \eqref{VaCond}--\eqref{VaCo} to other irreversible processes such as heat conduction, mass exchange, or chemical reactions, is the concept of \textit{thermodynamics displacement} introduced in \cite{GBYo2019}. By definition, the thermodynamic displacement $ \Lambda ^ \alpha (t)$ associated to an irreversible process $ \alpha $ is the primitive in time of the thermodynamic force $X^ \alpha (t)$ (see \cite{KoPr1998,dGMa1969}) of this process, i.e. $\dot \Lambda ^ \alpha (t)= X^ \alpha (t)$. For the case of heat conduction, the thermodynamic force is given by the temperature $T(t)$, hence the thermodynamic displacement is a variable $ \Gamma (t)$ with $\dot \Gamma (t)= T(t)$, called the \textit{thermal displacement}.
The introduction of $\Gamma (t)$ is accompanied by the introduction of an entropy variable $ \Sigma (t)$ whose meaning will be explained later. For the particular case below we will have $ \Sigma = S$, but this is not always true.

\subsubsection{Variational formulation for systems with internal heat exchanges} Consider a thermodynamic system consisting of subsystems, indexed by $k=1,...,N$, which are exchanging energy via the irreversible processes of heat conduction and mechanical friction. The Lagrangian function is given as $L:TQ \times \mathbb{R} ^N \rightarrow \mathbb{R} $, which depends on $(q, \dot q, S_1,...,S_N)$ with $S_k$, $k=1,...,N$, the entropy of subsystem $k$. The friction forces acting on each subsystems are denoted $F^{{\rm fr} (k)}:TQ \times \mathbb{R} ^N \rightarrow \mathbb{R} $. The fluxes associated with the heat exchange between subsystems $k$ and $l$ are denoted $J_{kl}$, $k\neq l$ and satisfy $J_{kl}=J_{lk}$. For the
construction of variational structures, it is convenient to define the flux $J_{kl}$ for $k = l$ as  $J_{kk}:= - \sum_{l\neq k} J_{kl}$, so that
\begin{equation}\label{div_0} 
\sum_{k=1}^NJ_{kl}=0\,.
\end{equation}
The extension of \eqref{VaCond}--\eqref{VaCo} to this case consists in the variational condition
\begin{equation}\label{VaCond_AB} 
\delta \int_0^T \left[ L(q, \dot q,S_1,...,S_N) + \dot \Gamma ^k(S_k- \Sigma _k) \right] {\rm d} t=0
\end{equation} 
subject to the constraint (phenomenological constraint)
\begin{equation}\label{PhCo_AB} 
\frac{\partial L}{\partial S_k} \dot \Sigma _k = \left\langle F^{{\rm fr}(k)} , \dot q \right\rangle + J_{kl} \dot \Gamma ^l, \quad \text{for $A=1,...,N$},
\end{equation} 
on the solution curve and the constraint (variational constraint)
\begin{equation}\label{VaCo_AB} 
\frac{\partial L}{\partial S_k} \delta  \Sigma _k = \left\langle F^{{\rm fr}(k)} , \delta  q \right\rangle + J_{kl} \delta  \Gamma ^l, \quad \text{for $k=1,...,N$}
\end{equation} 
on the variations. 

Application of the variational principle \eqref{VaCond_AB}--\eqref{VaCo_AB} yields
\begin{equation} 
\begin{aligned} 
\int_0^T & \left( - \frac{d}{dt} \pp{L}{\dot q } + \pp{L}{q} \right) \cdot \de q + \left( \pp{L}{S_k} + \dot \Gamma^k \right) \de S_k - 
\dot \Gamma^k \de \Sigma_k + \de \Gamma^k \left( \dot \Sigma_k - \dot S_k \right) \mbox{d} t 
\\ 
=& \int_0^T  \bigg( - \frac{d}{dt} \pp{L}{\dot q } + \pp{L}{q} + \sum_k F^{\rm fr(k)} \bigg) \cdot \de q +
\bigg( \dot \Sigma_k - \dot S_k - \frac{1}{\pp{L}{S_k}} \sum_l J_{kl} \bigg) \de \Gamma^k\\
& \hspace{3cm} 
+\de S_k  \bigg( \pp{L}{S_k} + \dot \Gamma^k \bigg) \,  \mbox{d} t=0 \,.
\label{deriv_thermo_variations} 
\end{aligned} 
\end{equation} 
From $ \delta \Gamma ^k$ and $ \delta S_k$ we conclude that 
\begin{equation}
\label{two_extra_conditions} 
\dot \Sigma _k= \dot S_k + \sum_{l=1}^N \frac{\dot \Gamma ^k}{ \frac{\partial L}{\partial S_k} }J_{lk} \qquad\text{and}\qquad \dot \Gamma ^k=- \frac{\partial L}{\partial S_k}, \qquad k=1,...,N\,. 
\end{equation}
From \eqref{div_0}, the first condition implies $\dot \Sigma _k= \dot S_k$ in this particular case. However, in general, $\dot S_k \neq \dot\Sigma_k$, which is the case for the continuum media we will consider later in more details. 

Since $T^k= - \frac{\partial L}{\partial S_k}$ is the temperature of subsystem $k$, the second condition in \eqref{two_extra_conditions} reads $ \dot \Gamma ^k= T^k$ which implies that $ \Gamma ^k$ is the thermal displacement for subsystem $k$. 

The coefficient multiplying $\de q$ in \eqref{deriv_thermo_variations} and the use of the first equation in \eqref{two_extra_conditions} in the thermodynamic condition \eqref{PhCo_AB} allows to close the system and write the equations of motion only in terms of variables $(q,\dot q, S_1, \ldots, S_N)$:
\begin{equation}\label{thermo_mech_AB}
\left\{ 
\begin{array}{l}
\displaystyle\vspace{0.2cm}\frac{d}{dt} \frac{\partial L}{\partial\dot q}- \frac{\partial L}{\partial q}= \sum_{k=1}^NF^{{\rm fr}(k)}\\
\displaystyle \frac{\partial L}{\partial S_k} \dot S_k = \left\langle F^{{\rm fr}(k)} , \dot q \right\rangle + \sum_{l=1}^N J_{kl} \left( \frac{\partial L}{\partial S_k}- \frac{\partial L}{\partial S_l} \right) , \quad k=1,...,N\,.
\end{array}\right.
\end{equation}

The total entropy $S=\sum_{k=1}^NS_k$ of the system evolves as
\[
\dot S= - \sum_{k=1}^N \frac{1}{T^k} \left\langle  F^{{\rm fr}(k)}, \dot q \right\rangle  +\sum_{k<l}^N J_{kl} \left( \frac{1}{T^l} - \frac{1}{T^k} \right) (T^l-T^k)\,. 
\]

\subsubsection{Two piston system}

Let us now consider the case of two pistons connected with a rigid rod. In spite of its deceptive simplicity, as it turns out, this model is an accurate discrete analogue of two media (solid and fluid) with an incompressibility constraint, which is the case of this paper. Consider two pistons $k=1,2$ each pressurising a tank filled with gas. The state of the gas and piston is described by the coordinate $x_k$ and entropy $S_k$. We assume a state equation $U_k=U_k(S_k,V_k,N_k)$ for each gas, where $U_k$ is the internal energy, $V_k$ the volume, and $N_k$ the number of moles. The Lagrangian of the system is
\begin{equation} 
\label{Lagr_twopistons} 
L(x_1, \dot x_1, x_2, \dot x_2, S_1, S_2)= \sum_{k=1}^2 \frac{1}{2} m_k\dot x_k ^2 - u_k(x_k, S_k)\,,
\end{equation} 
where the second term is defined from the internal energy of each gas as $u_k(x_k, S_k)= U_k(S_k, A_kx_k, N_k^0)$, with $A_k$ the area of the section of the pistons and $N_k^0$ the number of moles assumed to be constant. The distance between the bottom of the two pistons is $D$ while the length of the bar connecting the piston is $\ell$, hence we have the constraint
\begin{equation}\label{constraint_piston}
x_1+x_2+\ell=D,
\end{equation} 
see \cite{gay2017lagrangian}. We assume friction forces as $F^{{\rm fr} (k)}(x_k, \dot x_k, S_k) = - \lambda (x_k, S_k)\dot x_k$ and that the bar connecting the two pistons is heat conducting.

Application of the variational formulation \eqref{VaCond_AB}--\eqref{VaCo_AB} where $q=(x_1,x_2)$ and with the augmented Lagrangian
\begin{equation}\label{L_constraint} 
L(x_1, \dot x_1, x_2, \dot x_2, S_1, S_2) \quad \rightarrow \quad L(x_1, \dot x_1, x_2, \dot x_2, S_1, S_2) + \mu ( x_1+x_2+\ell -D)
\end{equation} 
yields the equations of motion as 
\begin{equation}\label{2_pistons}
\left\{ 
\begin{array}{l}
\displaystyle\vspace{0.2cm}m_1\ddot x_1 -p_1(x_1,S_1) A_1= - \lambda _1 \dot x_1+ \mu \\
\displaystyle\vspace{0.2cm}m_2\ddot x_2 -p_2(x_2,S_2) A_2= - \lambda _2 \dot x_2+ \mu \\
\displaystyle\vspace{0.2cm}T_1\dot S_1 = \lambda _1  \dot x_1 ^2 + J(T_2-T_1)\\
\displaystyle\vspace{0.2cm} T_2\dot S_2 = \lambda _2  \dot x_2 ^2 + J(T_1-T_2)\\
\displaystyle x_1+x_2+\ell=D\,,
\end{array}\right.\vspace{0.2cm}
\end{equation}
where we wrote $J=-J_{12}$. By eliminating the Lagrange multiplier and using the constraint, one gets the mechanical balance equation as
\begin{equation}\label{total_momentum} 
(m_1+m_2)\dot x =p_1 (x,S_1)A_1 - p_2(D-\ell - x,S_2) A_2 - ( \lambda _1+ \lambda _2)\dot x\,,
\end{equation} 
where we have written $x_1=x$.
We have chosen here to apply \eqref{VaCond_AB}--\eqref{VaCo_AB} to the Lagrangian augmented with the constraint \eqref{L_constraint} since it is a continuum version of this approach that is needed for the thermodynamics of porous media below when the fluid and/or the solid is incompressible. In the present case of the 2-piston, however, due to the simplicity of the constraint, one could directly apply \eqref{VaCond_AB}--\eqref{VaCo_AB} to the Lagrangian $L$ defined on the constrained space and deduce \eqref{total_momentum} without inserting the Lagrange multiplier, which is not the case for the porous media.

\begin{remark}[On  systems with heat-dependent constraints]
{\rm 
One can generalize the system discussed above to the case when constraints depend on entropy. For example, for the physical case of two connected pistons considered above, we can assume that the length $\ell$ of the rod connecting the pistons in  \eqref{constraint_piston} depends on temperatures of the pistons and hence on the entropies. That problem is technically quite challenging and intricate, and therefore will be considered in our future work. For the case of porous media we consider here, the incompressibility condition, which is an infinite-dimensional analogue of \eqref{constraint_piston}, does not depend on temperature. We refer to \cite{gay2017variational,ELGB2021} for applications of the variational formulation of thermodynamics for fluid models with constraints depending on the entropy.
}
\end{remark}

\subsection{Variational formulation for a heat conducting viscous fluid}\label{review_NSF}

We quickly review here the variational formulation for continuum mechanics introduced in  \cite{gay2017lagrangian2} that includes the irreversible processes of viscosity and heat condition. We start with the material description since it is in this description that the variational formulation is the simpler. The Eulerian variational formulation is then deduced from it.

\subsubsection{Lagrangian (material) description} In this description, the variational formulation is an extension of the Hamilton principle of continuum mechanics given in \eqref{HP} and a continuum version of the variational formulation of 
the nonequilibrium of finite dimensional systems with heat conduction recalled in \eqref{VaCond_AB}--\eqref{VaCo_AB}.

We take the same expression \eqref{General_L_continuum} for the Lagrangian  as in the reversible case, however now the material entropy density appears in the Lagrangian as an independent variable $S(t, \boldsymbol{X})$ rather than a parameter $S_0( \boldsymbol{X})$. Hence the Lagrangian is seen as a function $L: T \operatorname{Emb}( \mathcal{B} , \mathbb{R} ^3  ) \times \operatorname{Den}( \mathcal{B} ) \rightarrow \mathbb{R}$
\begin{equation}\label{General_L_continuum_thermo} 
L( \boldsymbol{\varphi }  , \dot{ \boldsymbol{\varphi }  }, S)= \int_ \mathcal{B} \left[ \frac{1}{2} \varrho _0 | \dot{\boldsymbol{\varphi } }|- \varrho_0  \,\mathcal{E} ( \mathbb{F}  , \varrho_0 , S, G_0) \right] {\rm d}^3 \boldsymbol{X}\,,
\end{equation}
with $ \operatorname{Den}( \mathcal{B} )$ the space of densities on $ \mathcal{B} $.
The continuum version of the variational formulation  \eqref{VaCond_AB}--\eqref{VaCo_AB} is given by the variational condition
\begin{equation}\label{VC_review} 
\delta \int_0^ T \Big[L (\boldsymbol{\varphi }  , \dot{ \boldsymbol{\varphi }  }, S) + \int_ \mathcal{B} (S- \Sigma ) \dot \Gamma \,{\rm d}^3 \boldsymbol{X}=0 \Big]{\rm d}t=0\, ,
\end{equation} 
subject to the phenomenological constraint
\begin{equation}\label{PC_review} 
\frac{\delta  L}{ \delta  S}\dot \Sigma = - \boldsymbol{P}  ^{\rm fr}
: \nabla \dot{\boldsymbol{\varphi }} + \boldsymbol{J}_S \cdot \nabla \dot{ \Gamma }
\end{equation} 
and with respect to variations $ \delta \boldsymbol{\varphi } $, $ \delta S$, $ \delta \Sigma $, $ \delta \Gamma $ subject to the variational constraint
\begin{equation}\label{VConstr_review} 
\frac{\delta  L}{ \delta  S}\delta \Sigma = - \boldsymbol{P}  ^{\rm fr}
: \nabla \delta \boldsymbol{\varphi } + \boldsymbol{J}_S \cdot \nabla \delta \Gamma\,.
\end{equation}

The tensor $\boldsymbol{P}  ^{\rm fr}$ is the Piola-Kirchhoff viscous stress tensor and $ \boldsymbol{J}_S$ is the entropy flux density in Lagrangian representation, see \cite{gay2017lagrangian2}. They are the Lagrangian objects corresponding to the more widely used Eulerian viscous stress tensor $ \boldsymbol{\sigma }^{\rm fr}$ and Eulerian entropy flux density $ \boldsymbol{j}_s$, see below.

As it will be clarified below, and in a similar way with their analogue for finite dimensional systems in \S\ref{subsec_example_piston}, the variable $ \Sigma(t ,\boldsymbol{X}) $ is the entropy generated by the irreversible processes and $ \Gamma (t ,\boldsymbol{X}) $ is the thermal displacement.

We note that one passes from the phenomenological constraint \eqref{PC_review} to the variational constraint \eqref{VConstr_review} by formally replacing each occurrence of a time derivative by a $ \delta $-variation, exactly as in the finite dimensional case considered earlier in \eqref{VaCond}--\eqref{VaCo}  and \eqref{VaCond_AB}--\eqref{VaCo_AB}. This has to be done for each of the irreversible processes considered; here $\boldsymbol{P}  ^{\rm fr}
: \nabla \dot{\boldsymbol{\varphi }} \rightarrow \boldsymbol{P}  ^{\rm fr}
: \nabla \delta \boldsymbol{\varphi }$ for viscosity and $\boldsymbol{J}_S \cdot \nabla \dot\Gamma \rightarrow \boldsymbol{J}_S \cdot \nabla \delta \Gamma$ for heat conduction.

\subsubsection{Eulerian (spatial) description} The spatial version of \eqref{VC_review}--\eqref{VConstr_review} gives the variational formulation
\begin{equation}\label{VC_review_spat} 
\delta \int_0^ T \Big[\ell (\boldsymbol{u}  , \rho  , s, b) + \int_ \mathcal{B} (s - \sigma ) D_t \gamma   \,{\rm d}^3 \boldsymbol{x} \Big]{\rm d}t=0\, ,
\end{equation} 
subject to the phenomenological constraint
\begin{equation}\label{PC_review_spat} 
\frac{\delta  \ell}{ \delta  s} \bar D_t \sigma = - \boldsymbol{\sigma }  ^{\rm fr}
: \nabla  \boldsymbol{u}  + \boldsymbol{j}_s \cdot \nabla D_t  \gamma 
\end{equation} 
and with respect to variations
\[
\delta \boldsymbol{u}= \partial _t \boldsymbol{\eta} + \boldsymbol{u} \cdot \nabla \boldsymbol{\eta} - \boldsymbol{\eta} \cdot \nabla \boldsymbol{u}, \quad \delta \rho  = - \operatorname{div}( \rho  \boldsymbol{\eta} ), \quad \delta b= - \pounds _ { \boldsymbol{\eta} }b, \quad \delta s, \quad \delta \sigma , \quad \delta \gamma 
\]
subject to the variational constraint
\begin{equation}\label{VConstr_review_spat} 
\frac{\delta \ell}{ \delta  s}\bar D_ \delta  \sigma = - \boldsymbol{ \sigma }  ^{\rm fr}
: \nabla \boldsymbol{\eta}  + \boldsymbol{j}_s \cdot \nabla D_ \delta  \gamma\,.
\end{equation}

In \eqref{VC_review_spat}--\eqref{VConstr_review_spat}, the variables $ \gamma (t, \boldsymbol{x}) $, $ \sigma  (t, \boldsymbol{x}) $, $ \boldsymbol{\sigma }^{\rm fr} (t, \boldsymbol{x}) $, and $ \boldsymbol{j}_s (t, \boldsymbol{x}) $ are the Eulerian quantities associated to $ \Gamma  (t, \boldsymbol{X}) $, $ \Sigma  (t, \boldsymbol{X}) $, $ \boldsymbol{P}^{\rm fr} (t, \boldsymbol{X}) $, and $ \boldsymbol{J}_S (t, \boldsymbol{X}) $, respectively. We have introduced the notations
\begin{equation}\label{D_t_notation}
\begin{aligned} 
&D_t f= \partial _t f + \boldsymbol{u} \cdot \nabla f &\qquad  & D_ \delta  f= \delta  f + \boldsymbol{\eta } \cdot \nabla f \\
&\bar D_t f = \partial _t f + \operatorname{div}(f \boldsymbol{u})  & \qquad  & \bar D_ \delta f = \delta  f + \operatorname{div}(f \boldsymbol{\eta })
\end{aligned}
\end{equation}
for the Lagrangian time derivative and Lagrangian variations of a scalar function and a density.

We assume no-slip boundary conditions $ \boldsymbol{u}=0$ on $ \partial \mathcal{B} $.
A direct application of the variational principle \eqref{VC_review_spat}--\eqref{VConstr_review_spat} yields the general equations of motion for a heat conducting viscous continuum in Eulerian coordinates as
\begin{equation}\label{reduced_EL_thermo} 
\left\{
\begin{array}{l}
\vspace{0.2cm}\displaystyle \partial_t \frac{\delta\ell}{\delta \boldsymbol{u} }+\pounds_{\boldsymbol{u}}\frac{\delta\ell}{\delta\boldsymbol{u}} =\rho\nabla \frac{\delta\ell}{\delta\rho}  +s\nabla \frac{\delta\ell}{\delta s}- \frac{\delta\ell}{\delta b}:\nabla b- 2\operatorname{div} \left( \frac{\delta\ell}{\delta b}\cdot b \right) + \operatorname{div} \boldsymbol{\sigma} ^{\rm fr}\\
\displaystyle\frac{\delta \ell}{\delta s}( \bar D_t s + \operatorname{div} \boldsymbol{j}_s) = - \boldsymbol{\sigma} ^{\rm fr}: \nabla \boldsymbol{u}- \boldsymbol{j}_s \cdot \nabla \frac{\delta \ell}{\delta s}
\end{array} \right. 
\end{equation} 
together with the conditions
\begin{equation}\label{two_extra_conditions_fluid} 
\bar D_t \sigma = \bar D_t s+ \operatorname{div} \boldsymbol{j}_s \qquad\text{and}\qquad \bar D_t \gamma = - \frac{\delta \ell}{\delta s}\,,
\end{equation} 
which are the continuum analogue to conditions \eqref{two_extra_conditions}. 
From these two conditions, $\bar D_t \sigma $ is interpreted as the total entropy generation rate density and $D_t \gamma $ is the temperature, hence $ \gamma$ is the thermal displacement, exactly as in \S\ref{subsec_example_piston}.

The equations for $ \rho $ and $b$ follow as earlier from \eqref{Eulerian_mass} and \eqref{Eulerian_Finger} as
\[
\partial _t \rho  + \operatorname{div}( \rho  \boldsymbol{u}  )=0 \quad\text{and}\quad  \partial _t b + \pounds _{ \boldsymbol{u}} b=0.
\]
By using the expression of the Lagrangian given in \eqref{Eulerian_Lagrangian}, the equations of motion \eqref{reduced_EL_thermo} give the following system of visco-elastic heat conducting continuum 
\begin{equation}\label{reduced_EL_thermo} 
\left\{
\begin{array}{l}
\vspace{0.2cm}\displaystyle
\rho  (\partial _t \boldsymbol{u} + \boldsymbol{u} \cdot \nabla \boldsymbol{u}) = - \nabla p + \operatorname{div} \boldsymbol{\sigma}_{\rm el}+ \operatorname{div} \boldsymbol{\sigma}^{\rm fr}\\
\displaystyle T(\bar D_t s + \operatorname{div} \boldsymbol{j}_s) =  \boldsymbol{\sigma} ^{\rm fr}: \nabla \boldsymbol{u}- \boldsymbol{j}_s \cdot \nabla T,
\end{array} \right. 
\end{equation} 
with $p$ the pressure, $T$ the temperature, and $ \boldsymbol{\sigma} _{\rm el}$ the elastic stress.
We refer to \cite{gay2017lagrangian,gay2017lagrangian2,GBYo2019,gay2017variational} for the statement of the variational formulation in both the material and spatial description as well as the detailed computations and several applications and extensions.

\subsection{Variational modeling of porous media thermodynamics}
\label{sec:Thermo_comp}

We derive the equations of motion for a porous media which takes into account of the irreversible processes of heat exchange as well as the friction forces and friction stresses between the two components. For simplicity, we do not include the heat conduction within each component but it can be easily included as done for a one component continua in \S\ref{review_NSF}.

We denote by $ \boldsymbol{f} _k$ and $ \boldsymbol{\sigma} _k$, $k=s,f$ the friction forces and stresses acting on the solid and the fluid, and by $J_{kl}$, $k,l \in \{s,f\}$ the fluxes associated to the heat exchange between the fluid and the solid. Exactly as in \S\ref{subsec_example_piston}, we assume $J _{kl} = J_{lk} $ for $k\neq l$, and $J_{kk}=- \sum_{l\neq k}J_{kl}$. In our case of two components, it suffices to know $J_{fs}$ and we have $J_{ss}=J_{ff}= - J_{sf}$.

We adapt the notation \eqref{D_t_notation} to  the case of two media as
\begin{equation}\label{D_t_notation_fs}
\begin{aligned} 
&D_t ^kf= \partial _t f + \boldsymbol{u}_k \cdot \nabla f & \qquad  & D_ \delta  ^kf= \delta  f + \boldsymbol{\eta }_k \cdot \nabla f \\
&\bar D_t^k f = \partial _t f + \operatorname{div}(f \boldsymbol{u}_k)  & \qquad  & \bar D_ \delta ^kf = \delta  f + \operatorname{div}(f \boldsymbol{\eta }_k)
\end{aligned}
 \qquad k=f,s\,.
\end{equation}

\subsubsection{Variational formulation and general equations} The variational formulation is found by writing a continuum version of \eqref{VaCond_AB}--\eqref{VaCo_AB} written in the Eulerian (spatial) setting of \eqref{VC_review_spat}--\eqref{VConstr_review_spat}. This immediately gives the variational formulation:
\begin{equation}\label{VP_Porousmedia}
\begin{aligned}
&\delta  \int_0^T \Big[\ell( \boldsymbol{u} _f,\boldsymbol{u}_s,\rho_f,\rho_s,s _f , s_s , b,\phi ) + \sum_{k=f,s}\int_ \mathcal{B} (s _k - \sigma _k  ) D_t ^k  \gamma _k \, {\rm d} ^3  \boldsymbol{x}\Big] {\rm d}t=0\, ,
\end{aligned}
\end{equation} 
subject to the \textit{phenomenological constraints}
\begin{equation}\label{PC_Porousmedia}
\begin{aligned} 
&\text{fluid:} \quad \frac{\delta  \ell }{\delta  s _f }\bar D_t^f \sigma  _f = \boldsymbol{f} _f \cdot \boldsymbol{u} _f - \boldsymbol{\sigma} _f : \nabla \boldsymbol{u} _f   + \sum_k J_{fk}D_t ^k \gamma_k\\
&\text{solid:} \quad  \frac{\delta  \ell }{\delta  s _s }\bar D_t ^s\sigma  _s = \boldsymbol{f}  _s \cdot \boldsymbol{u}_s   -  \boldsymbol{\sigma} _s: \nabla \boldsymbol{u}_s  + \sum_k J_{sk}D_t ^k \gamma_k\, ,
\end{aligned} 
\end{equation} 
and with respect to variations $ \delta \boldsymbol{u} _k  =\partial _t \boldsymbol{\eta} _k  + \boldsymbol{u}  _k  \cdot \nabla \boldsymbol{\eta} _k  - \boldsymbol{\eta} _k \cdot \nabla \boldsymbol{u}  _k $, $\delta \rho _k= - \operatorname{div}( \rho  _k \boldsymbol{\eta} _k  )$,  $\delta s_k $, $ \delta \sigma _k $, and $ \delta \gamma _k $, $k=f,s$, such that $ \boldsymbol{\eta} _k $, $ \delta \sigma _k $ and $ \delta \gamma _k $ satisfy the \textit{variational constraint}
\begin{equation}\label{VarC_Porousmedia}
\begin{aligned} 
&\text{fluid:} \quad \frac{ \delta \ell }{ \delta  s _f }\bar D_ \delta ^f  \sigma  _f = \boldsymbol{f} _f \cdot \boldsymbol{\eta}  _f - \boldsymbol{\sigma} _f : \nabla \boldsymbol{\eta}  _f   + \sum_k J_{fk}D_\delta^k \gamma_k\\
&\text{solid:} \quad \frac{\delta \ell  }{\delta  s _s }\bar D_\delta ^s \sigma  _s =  \boldsymbol{f} _s \cdot \boldsymbol{\eta}  _s  -  \boldsymbol{\sigma} _s : \nabla \boldsymbol{\eta}_s   + \sum_k J_{sk}D_\delta^k \gamma_k\, ,
\end{aligned} 
\end{equation}  
with $ \delta \gamma _k $, and $\boldsymbol{\eta} _k $ vanishing at $t=0,T$. The forces $\boldsymbol{f} _{f,s}$ and stresses $\boldsymbol{\sigma} _{f,s}$ are coming from friction and have to be postulated phenomenologically. As we shall see below, the variational principle allows to guide the search for exact forms for these expressions.

The variational formulation \eqref{VP_Porousmedia}--\eqref{VarC_Porousmedia}  yields the general system of equations
\begin{equation}\label{general_thermo} 
\!\!\!\left\{ 
\begin{array}{l}
\displaystyle\vspace{0.2cm} \partial_t \frac{ \delta \ell}{\delta  \boldsymbol{u} _f} + \pounds_{ \boldsymbol{u}_f}  \frac{ \delta \ell}{\delta  \boldsymbol{u} _f}= \rho_f \nabla \frac{\delta \ell }{\delta  \rho _f} + s_f \nabla \frac{\delta  \ell }{\delta  s_f} + \operatorname{div}\boldsymbol{\sigma}_f+ \boldsymbol{f}_f\\
\displaystyle\vspace{0.2cm} \partial_t\frac{ \delta \ell}{\delta  \boldsymbol{u} _s} + \pounds_{ \boldsymbol{u}_s}  \frac{ \delta \ell}{\delta  \boldsymbol{u} _s}= \rho_s \nabla \frac{\delta  \ell }{\delta  \rho _s} + s_s \nabla \frac{\delta  \ell }{\delta  s_s} - \frac{\delta \ell }{\delta b}:\nabla b+  \operatorname{div} \left( \boldsymbol{\sigma}_s-2 \frac{\delta \ell }{\delta b}\cdot b \right) + \boldsymbol{f} _s\\
\vspace{0.2cm}\displaystyle\partial_t \rho_f+ \operatorname{div}(\rho_f \boldsymbol{u}_f)=0,\qquad \partial_t \rho_s+ \operatorname{div}(\rho_s \boldsymbol{u}_s)=0,\qquad\partial_t b+ \pounds_{\boldsymbol{u}_s}b=0\\
\vspace{0.2cm}\displaystyle\frac{ \delta \ell}{\delta  s _f}  \bar D_t^f s_f = \boldsymbol{f} _f\cdot\boldsymbol{u}_f  - \boldsymbol{\sigma}_f:\nabla \boldsymbol{u} _f - J_{fs}\left(\frac{ \delta \ell}{\delta  s _s}-\frac{ \delta \ell}{\delta  s _f}\right)\\
\vspace{0.2cm}\displaystyle  \frac{ \delta \ell}{\delta  s _s}  \bar D_t^s s_s = \boldsymbol{f} _s\cdot\boldsymbol{u}_s 
 - \boldsymbol{\sigma}_s:\nabla\boldsymbol{u}_s  - J_{sf}\left( \frac{ \delta \ell}{\delta  s _f} - \frac{ \delta \ell}{\delta  s _s}\right)\\
\displaystyle\frac{ \delta \ell}{\delta  \phi}=0,
\end{array}
\right.
\end{equation} 
together with the conditions
\[
\bar D_t^k  \sigma _k =\bar D_t^k s _k \qquad\text{and}\qquad D _t ^k\gamma_k= - \frac{\delta \ell }{\delta   s _k },\quad k=f,s,
\]
which have allowed to eliminate $ \sigma _k $ and $ \gamma _k $ in the final equations, in a similar way with \eqref{two_extra_conditions} and \eqref{two_extra_conditions_fluid}.

\subsubsection{Porous media thermodynamics with compressible components}

The goal of this section is to derive the general form for the stress tensors based on the thermodynamically consistent equations derived in this article. As it turns out, particularly simple general expressions for the viscous stress follow from the thermodynamic consistency and isotropy considerations. 
From the general background in the continuum mechanics applied to the theory of mixtures \cite[p.216]{atkin1976continuum}, we know that $\boldsymbol{\sigma}_f+\boldsymbol{\sigma}_s$ must be a symmetric tensor.  
A particular choice, most natural in the theory of porous media, is $\boldsymbol{\sigma}_f+\boldsymbol{\sigma}_s=0$. This choice arises from thinking of the quantities describing a porous media to be defined in a volume that is much smaller than the scale of the system, but much larger than the size of the pores. In that volume, averages over randomly positioned pores will average out the internal fluctuations of the stress tensor inside the fluid and solid, from pore to pore. We stress that it is a \emph{physical} assumption, and while it is likely to be valid with a high accuracy for many materials, it is likely that this assumption will not be satisfied for other materials having a different internal structure. Thus, we also present the case where $\boldsymbol{\sigma}_f+\boldsymbol{\sigma}_s \neq 0$ later in this section. Thus, for now, we assume $\boldsymbol{f}  _f =- \boldsymbol{f} _s =:  \boldsymbol{f}$ and $\boldsymbol{\sigma}_f=-\boldsymbol{\sigma}_s=:\boldsymbol{\sigma}$. 


By using the Lagrangian $\ell$ given in \eqref{Lagr_def}, we get the system
\begin{equation}\label{thermodynamics_porousmedia}
\left\{
\begin{array}{l}
\vspace{0.2cm}\displaystyle \rho_f (\partial_t \boldsymbol{u}_f+ \boldsymbol{u}_f\cdot \nabla \boldsymbol{u}_f) = - \phi \nabla p + \operatorname{div}\boldsymbol{\sigma}+ \boldsymbol{f}  \\
\vspace{0.2cm}\displaystyle\rho_s( \partial_t \boldsymbol{u}_s+ \boldsymbol{u}_s\cdot \nabla \boldsymbol{u}_s) = - (1-\phi) \nabla p  +  \operatorname{div} \left( \boldsymbol{\sigma} _{\rm el}  -  \boldsymbol{\sigma} \right) -  \boldsymbol{f}\\
\vspace{0.2cm}\displaystyle\partial_t \rho_f+ \operatorname{div}(\rho_f \boldsymbol{u}_f)=0,\qquad \partial_t \rho_s+ \operatorname{div}(\rho_s \boldsymbol{u}_s)=0,\qquad\partial_t b+ \pounds_{\boldsymbol{u}_s}b=0\\
\vspace{0.2cm}\displaystyle T_f  \bar D_t^f s_f =- \boldsymbol{f} \cdot\boldsymbol{u}_f  + \boldsymbol{\sigma}:\nabla\boldsymbol{u}_f - J_{fs}\left(T_s-T_f\right)\\
\displaystyle T_s  \bar D_t^s s_s =  \boldsymbol{f} \cdot\boldsymbol{u}_s    -\boldsymbol{\sigma}:\nabla\boldsymbol{u}_s  - J _{sf}\left(T_f-T_s\right) 
\end{array}
\right.
\end{equation}
together with the condition \eqref{p_compressible}, which defines the pressure $p$ appearing in the momentum equation. We defined the temperatures of fluid and solid as
\[
T_k= -\frac{\delta  \ell}{ \delta  s _k} = \frac{\partial e _k}{\partial \eta _k } , \qquad k=s,f.
\]
The total entropy equation is computed as
\begin{equation} 
\label{entropy_condition}
\bar D_t^f s_f +  \bar D_t^s s_s = \boldsymbol{f}  \cdot \Big(  \frac{ \boldsymbol{u} _s }{T_s}- \frac{ \boldsymbol{u} _f }{T_f}  \Big)  +
\boldsymbol{\sigma}: \left( \frac{1}{T_f}  \nabla \bu_f-\frac{1}{T_s}  \nabla \bu_s\right) - \Big( \frac{1}{T_f} - \frac{1}{T_s} \Big) J (T_f-T_s),
\end{equation} 
where we defined $J:=- J_{sf}= - J_{fs}$. 
This gives conditions on the expressions of $ \boldsymbol{f}  $, $ \boldsymbol{\sigma} _f $, and $J$, to satisfy the second law of thermodynamics. For instance \eqref{entropy_condition} suggests the expressions
\begin{equation}\label{phenomenology_1} 
\boldsymbol{f}  = \mathbb{K} \Big(  \frac{ \boldsymbol{u} _s }{T_s}- \frac{ \boldsymbol{u} _f }{T_f}  \Big) \qquad\text{and}\qquad J\geq 0,
\end{equation} 
where $ \mathbb{K}$ is a symmetric positive operator that can in general depend on the variables $ \rho  _f$, $\rho  _s$, $s_f$, $s_s$, $b$, $\phi$, similarly for $J$. 

The condition \eqref{phenomenology_1} is just the Darcy law of friction in porous media with constituent having different temperatures. The term involving $\boldsymbol{\sigma}$ is more interesting and leads to new considerations on the friction terms in porous media. 
For porous media, one commonly assumes that the force acting on the fluid and solid is given by the expression \eqref{phenomenology_1} coming from computing the friction force of the relative motion of the fluid with respect to the pores in the lubrication approximation. However, there is another part of friction forces that is commonly neglected, related to the relative motion of the pores with respect to each other, and resulting in corresponding induced motion of the fluid.
Suppose there is a local shearing motion of the solid, \emph{i.e.}, a non-trivial value of $\nabla \boldsymbol{u} _s$. The boundaries of an individual pore will move at different speed, inducing the motion of the fluid inside the pore that does not contribute to the net motion (\emph{i.e.}, it averages to the average fluid speed $\bu_f$). However, even though that microscopic motion of the fluid does not contribute to the net macroscopic velocity $\bu_f$, the microscopic motion does contribute to dissipation. A natural relation that fulfills  \eqref{entropy_condition} is
\begin{equation} 
\label{sigma_expression} 
\boldsymbol{\sigma}=\mathbb{C} : \left( \frac{1}{T_f}  \nabla \bu_f-\frac{1}{T_s}  \nabla \bu_s\right)\,,
\end{equation} 
where $\mathbb{C}$ is a positive $(2,2)$ type tensor in the sense that for all matrices $\mathbb{F}$ with components $F^i_j$ we have 
\begin{equation} 
\label{pos_def_C}
\mathbb{F}: \mathbb{C}: \mathbb{F} = C^{jp}_{iq} F^i_j  F^q_p \geq 0  \quad (\mbox{summation over repeated indices}) \, . 
\end{equation} 
\rem{ 
For example, borrowing from the linear elasticity, for isotropic uniform media we can assume that 
\begin{equation} 
\boldsymbol{\sigma} = 2 \mu  \mathbb{F} +  \lambda {\rm Id}\,   {\rm tr} \mathbb{F}\, , \quad \mathbb{F}:= \frac{1}{2} \left( \frac{1}{T_f}  \left( \nabla \bu_f+ \nabla \bu_f^T\right)-\frac{1}{T_s}  \left( \nabla \bu_s+ \nabla \bu_s\right)^T\right)  \, , 
\label{C_tensor_example} 
\end{equation} 
similar to the Lam\'{e} coefficients for an elasticity tensor for an isotropic elastic media. Here, ${\rm Id}$ is a $3 \times 3$ identity matrix and the coefficients $\lambda$ and $\mu$ satisfy
\begin{equation} 
\label{condition_coeff_Lame} 
\mu \geq 0 \, , \quad \mbox{and} \quad \lambda + \frac{2}{3} \mu \geq 0 \, . 
\end{equation} 
More generally, we 
} 

\color{black} 

Borrowing ideas from elasticity theory and fluid mechanics regarding the most general form of isotropic uniform tensors in 3D space, we can choose $\mathbb{C}$ using the isotropic+deviatoric+skew-symmeric decomposition as 
\begin{equation} 
\label{C_tensor_example_2} 
\begin{aligned} 
\boldsymbol{\sigma} &=  \zeta  \Big( \frac{1}{T_f} \operatorname{div} \mathbf{u} _f -  \frac{1}{T_s} \operatorname{div} \mathbf{u} _s \Big) {\rm Id} + 2\mu  \Big( \frac{1}{T_f}   \mathbb{F} _f ^{\,\circ} - \frac{1}{T_s} \mathbb{F}  _s ^{\,\circ} \Big) + \nu \Big( \frac{1}{T_f}   \mathbb{A}_f -\frac{1}{T_s}   \mathbb{A} _s  \Big) \, , 
\end{aligned} 
\end{equation} 
where $\mathbb{F}_{f,s}$ and $\mathbb{A}_{f,s}$ are the symmetric and antisymmetric parts of the velocity gradients,  and $\mathbb{F} _{f,s}^{\,\circ}$ is the deviatoric (traceless) part of the symmetrized velocity: 
\begin{equation} 
\mathbb{F}_{f,s} = \frac{1}{2} \big( \nabla \bu_{f,s} +\nabla  \bu_{f,s}^\mathsf{T} \big) \, , \quad  \mathbb{A}_{f,s} = \frac{1}{2} \big( \nabla \bu_{f,s} -\nabla  \bu_{f,s}^\mathsf{T} \big) \, , \quad 
\mathbb{F}_{f,s}^{\,\circ}    = \mathbb{F}_{f,s}- \frac{1}{3} {\rm Id} \, \operatorname{tr} \mathbb{F}_{f,s} \, . 
\label{vel_grad} 
\end{equation} 
The conditions on the parameters $\mu , \zeta , \nu$ are
\begin{equation} 
\mu \geq 0 \, ,   \quad \zeta  \geq 0 \, , \quad \nu \geq 0 \, . 
\label{param_cond} 
\end{equation} 
These equations generalize the Darcy-Brinkman laws of porous media \cite{brinkman1949calculation,brinkman1949permeability,brinkman1952viscosity}, see also \cite{kannan2008flow}, which in our notation and the absence of external forces will be written as  
\begin{equation} 
-  \nabla  p + \mu_f \nabla^2 \bu_f - \alpha \bu_f  =0 \, , 
\label{Darcy_Brinkman}
\end{equation} 
with $\mu_f$ being the effective dynamic viscosity of the fluid in porous media.  As far as we are aware, the conditions on the dissipative stress terms we have derived here have not been used in the previous works on the porous media. Equations \eqref{thermodynamics_porousmedia} provide a consistent description of the moving solid and fluid media with thermodynamics in a single system. The derivation presented here generalizes the thermodynamics foundation of mixture theory  presented in \cite{srinivasan2014thermodynamic} for non-moving porous media.

A more general case which includes viscosity in both components is the case in which the total stress $ \boldsymbol{\sigma} _f+ \boldsymbol{\sigma} _s$ does not vanish but is symmetric. A general relation is 
\[
\begin{bmatrix}
\vspace{0.2cm}\boldsymbol{\sigma} _f\\
\boldsymbol{\sigma} _s
\end{bmatrix}= 
\mathbb{L}\begin{bmatrix}
\frac{\nabla \boldsymbol{u}_f}{T_f} \\
\frac{\nabla \boldsymbol{u}_s}{T_s}
\end{bmatrix}, \qquad \mathbb{L}=\begin{bmatrix}
\vspace{0.2cm}\mathbb{A} & \mathbb{B}\\
\mathbb{C} & \mathbb{D}
\end{bmatrix},
\]
where the (2,2) type tensors $\mathbb{A}, \mathbb{B}, \mathbb{C}, \mathbb{D}$ are such that $ \mathbb{L}$ is positive and such that $ \boldsymbol{\sigma} _f+ \boldsymbol{\sigma} _s$ is symmetric. The natural generalization of \eqref{vel_grad} in this case is found as
\begin{equation} 
\label{sigma_fs_general} 
\begin{aligned} 
\boldsymbol{\sigma} _f&= \zeta _{ff} \frac{\operatorname{div} \boldsymbol{u}_f}{T_f}{\rm Id} +   2 \mu _{ff} \frac{\mathbb{F} ^{\, \circ }_f}{T_f}  + \zeta _{fs} \frac{\operatorname{div} \boldsymbol{u}_s}{T_s}{\rm Id} +   2 \mu _{fs} \frac{\mathbb{F} ^{\, \circ }_s}{T_s}  +   \nu \left( \frac{1}{T_f}   \mathbb{A}_f -\frac{1}{T_s}   \mathbb{A} _s  \right)\\
\boldsymbol{\sigma} _s&= \zeta _{sf} \frac{\operatorname{div} \boldsymbol{u}_f}{T_f}{\rm Id} +   2 \mu _{sf} \frac{\mathbb{F} ^{\, \circ }_f}{T_f}  + \zeta _{ss} \frac{\operatorname{div} \boldsymbol{u}_s}{T_s}{\rm Id} +   2 \mu _{ss} \frac{\mathbb{F} ^{\, \circ }_s}{T_s}  -  \nu \left( \frac{1}{T_f}   \mathbb{A}_f -\frac{1}{T_s}   \mathbb{A} _s  \right)\,,
\end{aligned} 
\end{equation} 
with conditions on parameters 
\begin{equation} 
\label{mu_zeta_cond}
\mu _{ff}\geq 0, \quad 4 \mu _{ff} \mu _{ss} \geq ( \mu _{fs}+ \mu _{sf}) ^2 , \quad \zeta  _{ff}\geq 0, \quad 4 \zeta _{ff} \zeta _{ss} \geq ( \zeta _{fs}+ \zeta _{sf}) ^2 , \quad \nu \geq 0\,. 
\end{equation} 
Equations \eqref{sigma_fs_general} with conditions \eqref{mu_zeta_cond} define the most general, thermodynamically consistent choice for the stress tensors. The nature of the parameter $\mu_{kl}$, $\zeta_{kl}$, and $ \nu $ will have to be determined experimentally.

\subsubsection{Energy balances} The balance of energy for each component and for the total system are useful to understand the impact of the irreversible processes. This uses the following result.

\begin{lemma}\label{Var_Energy} For an internal energy function of the form
\[
e_s \left( \frac{\rho  _s}{1- \phi }, \frac{s_s}{ \rho  _s}, b \right)\, ,
\]
we have
\begin{align*} 
\bar D_t^s ( \rho  _s e_s) 
= \left( e_s + \frac{\partial e_s}{\partial \bar \rho  _s}\bar \rho  _s - \frac{\partial e_s}{\partial \eta _s } \eta _s \right) \bar D^s_t \rho  _s  - \bar \rho  _s^2 \frac{\partial e_s}{\partial \bar \rho  _s}  \bar D^s_t (1- \phi ) + \rho  _s \frac{\partial e_s}{\partial b}: D_t b+T_s    \bar D^s_ts_s\, .
\end{align*}
\end{lemma}

By using the equations \eqref{thermodynamics_porousmedia}, we get the balance of internal energies for the fluid and the solid as
\begin{align*} 
\bar D^f_t( \rho  _f e_f) &= -p \bar D^f_t \phi  + \boldsymbol{\sigma }_f: \nabla \boldsymbol{u}_f- \boldsymbol{f}_f \cdot \boldsymbol{u}_f + J(T_s-T_f)\\
\bar D^s_t( \rho  _s e_s) &= -p \bar D^s_t( 1- \phi )  + \boldsymbol{\sigma }_{\rm el}: \nabla \boldsymbol{u}_s + \boldsymbol{\sigma }_s: \nabla \boldsymbol{u}_s - \boldsymbol{f}_s \cdot \boldsymbol{u}_s + J(T_f-T_s)\,.
\end{align*}
This shows that the variations of internal energies of each component is due to the friction forces and stresses, to the heat exchanges between the components and also to the changes in volume fraction.
The balance of total energies for the fluid and solid are deduced as
\begin{align*} 
\bar D^f_t\left(  \frac{1}{2} \rho  _f | \boldsymbol{u}_f| ^2 +   \rho  _f e_f\right) &=  \operatorname{div}\Big( (- \phi \, p \operatorname{Id}  + \boldsymbol{\sigma} _f) \cdot \boldsymbol{u}_f\Big) - p \partial _t \phi + J(T_s-T_f)\\
\bar D^s_t\left(  \frac{1}{2} \rho  _s | \boldsymbol{u}_s| ^2 +   \rho  _s e_s\right) &=  \operatorname{div}\Big(  (- (1-\phi ) p \operatorname{Id} + \boldsymbol{\sigma }_{\rm el} + \boldsymbol{\sigma} _s )\cdot \boldsymbol{u}_s\Big) + p \partial _t \phi + J(T_f-T_s)
\end{align*} 
which shows the density of power $p \partial _t \phi + J(T_f-T_s)$ exchanged by the solid and fluid, which cancels in the total energy balance which takes the form
\begin{align*} 
&\bar D^f_t\left(  \frac{1}{2} \rho  _f | \boldsymbol{u}_f| ^2 +   \rho  _f e_f\right) + \bar D^s_t\left(  \frac{1}{2} \rho  _s | \boldsymbol{u}_s| ^2 +   \rho  _s e_s\right)\\
& \qquad \qquad \qquad  =\operatorname{div}\Big( (- \phi \, p \operatorname{Id}  + \boldsymbol{\sigma} _f) \cdot \boldsymbol{u}_f +  (- (1-\phi ) p \operatorname{Id} + \boldsymbol{\sigma }_{\rm el} + \boldsymbol{\sigma} _s )\cdot \boldsymbol{u}_s\Big) \,.
\end{align*}

\subsection{Incompressibility and thermodynamics} 
\label{sec:incompr_fluid_thermodynamics}

For a single viscous and heat conducting fluid in the \textit{incompressible case}, i.e. for the system \eqref{reduced_EL_thermo} with the condition $ \operatorname{div} \boldsymbol{u} =0$, the thermodynamics plays a very minor role if the viscous stress does not depend on the temperature. Indeed, due to the incompressibility constraint, the pressure is found from a Poisson equation and is not related to the state equation of the fluid. Also, the entropy equation decouples from the momentum and mass density equation.
We will see that for the case for a porous media with an incompressible fluid the state equation of the fluid does play a role in the equations and the entropy equation is always fully coupled to the other evolution equations.

\subsubsection{Compressible solid with incompressible fluid}

The incompressibility constraint can easily be included in the variational formulation of nonequilibrium thermodynamics \eqref{VP_Porousmedia}--\eqref{VarC_Porousmedia} by doing the replacement
\begin{equation}
\begin{aligned} 
&\ell( \boldsymbol{u} _f, \boldsymbol{u} _s,\rho_f,\rho_s, s_f, s _s, b,\phi) \\
& \qquad \qquad \longrightarrow \quad  \ell( \boldsymbol{u} _f, \boldsymbol{u} _s,\rho_f,\rho_s, s_f, s _s, b,\phi)
 +  \int_\mathcal{B} \mu  \Big( \phi - (\phi^0 \circ \boldsymbol{\varphi} _f ^{-1})J\boldsymbol{\varphi} _f^{-1}\Big){\rm d} ^3\boldsymbol{x} 
\end{aligned} 
\label{Lagr_compressible_incompressible_thermodynamics} 
\end{equation} 
in the action functional \eqref{VP_Porousmedia} and considering $ \mu $ as an additional variable with free variations.
Proceeding analogously to the case above, we get the system \eqref{thermodynamics_porousmedia} where the pore volume fraction $ \phi $ is now governed by the additional equation \eqref{phi_equation_f} and with the pressure condition given \eqref{p_compressible_incompressible}.

As in the reversible case this condition gives the expression of the Lagrange multiplier in terms of the other variables.

As opposed to the case of a single incompressible fluid, from the thermodynamic consistency discussed earlier in \eqref{phenomenology_1} and \eqref{sigma_expression}--\eqref{C_tensor_example_2}, the friction forces and stresses necessarily depend on the temperatures $T_f$ and $T_s$. Hence, while the pressure in the fluid is not related to the fluid state equation, the entropy equation does not decouple and the system of equations depends on the fluid state equation.

For an incompressible fluid, the internal and total energy equations of the fluid do not depend on the rate of change the pore volume fraction as one gets
\begin{equation}\label{energy_equation_incompfluid}
\begin{aligned} 
\bar D^f_t( \rho  _f e_f) &= \boldsymbol{\sigma }_f: \nabla \boldsymbol{u}_f- \boldsymbol{f}_f \cdot \boldsymbol{u}_f + J(T_s-T_f)\\
\bar D^f_t\Big(  \frac{1}{2} \rho  _f | \boldsymbol{u}_f| ^2 +   \rho  _f e_f\Big) &= - \phi \nabla  p\cdot \boldsymbol{u}_f + \operatorname{div} ( \boldsymbol{\sigma} _f  \cdot \boldsymbol{u}_f)  + J(T_s-T_f)\,.
\end{aligned}
\end{equation}

\subsubsection{Incompressible fluid and incompressible solid} Analogously to \S\ref{sec:double_incompr}, we introduce two Lagrange multipliers $\mu_f$ and $\mu_s$ enforcing the incompressibility conditions of fluid and solid, respectively. Hence we use the variational formulation \eqref{VP_Porousmedia}--\eqref{VarC_Porousmedia} wih the replacement
\begin{equation}\label{Lagr_incompressible_incompressible_thermodynamics} 
\begin{aligned} 
& \ell( \boldsymbol{u} _f,\boldsymbol{u}_s,\rho_f,\rho_s,s _f , s_s , b,\phi) \quad \longrightarrow \\
& \qquad\ell( \boldsymbol{u} _f, \boldsymbol{u} _s,\rho_f,\rho_s,s _f , s_s , b,\phi)   +  \int_\mathcal{B} \mu_f  \Big( \phi - (\phi^0 \circ \boldsymbol{\varphi} _f ^{-1})J \boldsymbol{\varphi}_f ^{-1} \Big){\rm d} ^3   \boldsymbol{x} 
\\
&\hspace{5.15cm}+
\int_\mathcal{B} \mu_s  \Big( (1-\phi) - \big( (1-\phi^0) \circ \boldsymbol{\varphi} _s ^{-1}\big)J \boldsymbol{\varphi}_s^{-1}\Big){\rm d} ^3 \boldsymbol{x} \,.
\end{aligned}
\end{equation} 
This yields the system \eqref{thermodynamics_porousmedia} together with the two equations \eqref{phi_equations_incom_incomp} and the pressure condition \eqref{p_incompressible_incompressible}. Here, similarly to previous considerations, the dissipative forces and stresses follow the thermodynamically consistency forms suggested by \eqref{phenomenology_1} and \eqref{sigma_expression}--\eqref{C_tensor_example_2}. In addition to the fluid internal energy in \eqref{Lagr_incompressible_incompressible_thermodynamics} the solid internal energy equations also do not depend on the time variation of the pore volume fraction:
\[
\begin{aligned} 
\bar D^s_t( \rho  _s e_s) &= ( \boldsymbol{\sigma} _{\rm el}+ \boldsymbol{\sigma }_s): \nabla \boldsymbol{u}_s- \boldsymbol{f}_s \cdot \boldsymbol{u}_s + J(T_f-T_s)\\
\bar D^s_t\Big(  \frac{1}{2} \rho  _s | \boldsymbol{u}_s| ^2 +   \rho  _s e_s\Big) &= - (1-\phi)\nabla  p\cdot \boldsymbol{u}_s + \operatorname{div} ( ( \boldsymbol{\sigma} _{\rm el} + \boldsymbol{\sigma} _f) \cdot \boldsymbol{u}_s)  + J(T_f-T_s)\,.
\end{aligned}
\]
Thus, the variational thermodynamics approach developed here allows a unified approach to both compressible and incompressible cases, with physical meaning assigned to the Lagrange multipliers in terms of pressures. 

\begin{remark}[On the analogy between discrete and continuous models]
{\rm We hope that by now, the reader can connect the analogy between the two-piston system described in the previous section and the porous media with thermodynamics and incompressibility constraints. We have the following analogies between the variational formulations of the discrete and continuum cases:

\begin{equation} 
\begin{array}{c c c}
\hline \vspace{-0.2cm}\\
\vspace{0.2cm}\mbox{Discrete} & \qquad \qquad & \mbox{Continuous} \\
\hline \vspace{-0.3cm}\\
\vspace{0.1cm}\displaystyle L = \sum_{k=1}^2 \frac{1}{2} m_k\dot x_k ^2 - u_k(x_k, S_k)\;\;  & \qquad \qquad &\displaystyle \ell = \sum_k \int_ \mathcal{B} \Big[ \frac{1}{2}\rho_k|\boldsymbol{u}_k|^2 - \rho_k  e_k\Big] {\rm d}^3 \boldsymbol{x} \\
\hline \vspace{-0.1cm}\\
\vspace{0.3cm}\displaystyle\left\langle F^{{\rm fr} k}, \dot x _k \right\rangle &\qquad \qquad &\boldsymbol{f}  _k \cdot \boldsymbol{u}_k - \boldsymbol{\sigma} _k: \nabla \boldsymbol{u}_k\\
\hline \vspace{-0.2cm}\\
\vspace{0.2cm}\displaystyle\sum_k J_{kl} \dot \Gamma ^l &\qquad \qquad &\displaystyle \sum_k J_{kl} D_t^l \gamma _l\\
\hline \vspace{-0.2cm}\\
\vspace{0.2cm}\mu (x_1 + x_2 + \ell - D) &\qquad \qquad &\displaystyle\int_\mathcal{B} \mu  \Big( \phi - (\phi^0 \circ \boldsymbol{\varphi} _f ^{-1})J\boldsymbol{\varphi} _f^{-1}\Big){\rm d} ^3\boldsymbol{x} \\
\hline 
\end{array} 
\end{equation} 
}
\end{remark}
Note that the problem of the piston movement has been an extremely subtle system in nonequilibrium thermodynamics, see e.g. \cite{Gr1999,kozlov2005piston}. The porous media is an infinite-dimensional analogy of the connected piston problem. Thus, we believe that the mathematical analysis of the equations derived here is likely to be highly non-trivial.

\section{Conclusions} 
In this paper, we have developed a novel variational theory for the thermodynamics of a two-phase porous media consisting of an elastic porous matrix completely filled with a fluid. We have included the effect of the irreversible processes due to the friction forces and stresses as well as possible heat exchanges between the two media. We have considered the cases when both the fluid and the solid can be compressible or incompressible. The equations we derived are valid for arbitrary Lagrangians and equations of state and are thus applicable to a wide range of problems. Our equations are especially important for applications exhibiting large deformations and strong dependence on temperatures, such as geophysical flows and nuclear industry. In a future work, we are going to study the practical applications of the derived equations, for instance by analysing how the propagation of elastic waves in the porous media is affected by thermodynamics effects.

This work has shown that the variational formulation of thermodynamics provides an efficient modeling tool for continua, well adapted to also cover the case of systems with constraints and degenerate Lagrangians.
Our approach has shown that incompressibility constraints, which are \textit{not} simply given here by a divergence free condition, can be easily included in the variational setting, allowing for a unified treatment of all the compressible/incompressible cases with irreversible processes. Other types of constraints can also be treated using this setting, as illustrated in \cite{gay2017variational,ELGB2021}. Also, due to the dependence of the model on the fluid fraction $ \phi $ only, and not on its rate $ \partial _t \phi $, all the Lagrangians appearing in this paper are degenerate, which do not present any particular difficulty in the variational setting presented here. These are advantages of the variational formulation compared to other formalisms for thermodynamics, such as the ones based on appropriate modifications of Poisson brackets \cite{Ka1984,Gr1984,Mo1984a,EdBe1991a,GrOt1997}, which may not behave well in the presence of constraints and degeneracies.


\section*{Acknowledgments}
We acknowledge fruitful discussions with  C. Doering, D. D. Holm, A. Ibraguimov, T. S. Ratiu, and H. Yoshimura.







\providecommand{\url}[1]{\texttt{#1}}
\providecommand{\urlprefix}{}
\providecommand{\foreignlanguage}[2]{#2}
\providecommand{\Capitalize}[1]{\uppercase{#1}}
\providecommand{\capitalize}[1]{\expandafter\Capitalize#1}
\providecommand{\bibliographycite}[1]{\cite{#1}}
\providecommand{\bbland}{and}
\providecommand{\bblchap}{chap.}
\providecommand{\bblchapter}{chapter}
\providecommand{\bbletal}{et~al.}
\providecommand{\bbleditors}{editors}
\providecommand{\bbleds}{eds: }
\providecommand{\bbleditor}{editor}
\providecommand{\bbled}{ed.}
\providecommand{\bbledition}{edition}
\providecommand{\bbledn}{ed.}
\providecommand{\bbleidp}{page}
\providecommand{\bbleidpp}{pages}
\providecommand{\bblerratum}{erratum}
\providecommand{\bblin}{in}
\providecommand{\bblmthesis}{Master's thesis}
\providecommand{\bblno}{no.}
\providecommand{\bblnumber}{number}
\providecommand{\bblof}{of}
\providecommand{\bblpage}{page}
\providecommand{\bblpages}{pages}
\providecommand{\bblp}{p}
\providecommand{\bblphdthesis}{Ph.D. thesis}
\providecommand{\bblpp}{pp}
\providecommand{\bbltechrep}{}
\providecommand{\bbltechreport}{Technical Report}
\providecommand{\bblvolume}{volume}
\providecommand{\bblvol}{Vol.}
\providecommand{\bbljan}{January}
\providecommand{\bblfeb}{February}
\providecommand{\bblmar}{March}
\providecommand{\bblapr}{April}
\providecommand{\bblmay}{May}
\providecommand{\bbljun}{June}
\providecommand{\bbljul}{July}
\providecommand{\bblaug}{August}
\providecommand{\bblsep}{September}
\providecommand{\bbloct}{October}
\providecommand{\bblnov}{November}
\providecommand{\bbldec}{December}
\providecommand{\bblfirst}{First}
\providecommand{\bblfirsto}{1st}
\providecommand{\bblsecond}{Second}
\providecommand{\bblsecondo}{2nd}
\providecommand{\bblthird}{Third}
\providecommand{\bblthirdo}{3rd}
\providecommand{\bblfourth}{Fourth}
\providecommand{\bblfourtho}{4th}
\providecommand{\bblfifth}{Fifth}
\providecommand{\bblfiftho}{5th}
\providecommand{\bblst}{st}
\providecommand{\bblnd}{nd}
\providecommand{\bblrd}{rd}
\providecommand{\bblth}{th}



\end{document}